\documentclass[aps, prd, amsmath, amssymb, floats, floatfix, groupedaddress, nofootinbib, twocolumn, showpacs,reprint,superscriptaddress]{revtex4-2}

\usepackage{threeparttable}
\usepackage{tablefootnote}
\usepackage{array}
\usepackage{makecell}
\usepackage{float}
\usepackage{latexsym}
\usepackage{graphicx}
\usepackage{amsmath}
\usepackage{amssymb}
\usepackage{amsfonts}
\usepackage{times}
\usepackage{xspace} 
\usepackage[usenames]{color}
\usepackage{dcolumn}
\usepackage{bm}
\usepackage{mathrsfs}
\usepackage{tikz}
\usepackage{appendix}
\usepackage{breqn}
\usepackage[]{units}      
\usepackage{enumitem}

\usepackage{subfigure}		
\usepackage[normalem]{ulem} 

\usepackage[tracking=true]{microtype}
\SetTracking{}{500}
\SetTracking{encoding={*}, shape=sc}{40}

\usepackage[normalem]{ulem} 

\makeatletter
\let\cat@comma@active\@empty
\makeatother

\begin{document} 
 
\title{Self-force on a static scalar charge in traversable wormholes}

\author{Jerome P. Mecca}
\email{jerome.mecca@unp.edu.ph}
\affiliation{University of Northern Philippines, Vigan City 2700, Philippines}
\affiliation{National Institute of Physics, University of the Philippines, Diliman, Quezon City 1101, Philippines}

\author{Ian Vega}
\email{ivega@nip.upd.edu.ph}
\affiliation{National Institute of Physics, University of the Philippines, Diliman, Quezon City 1101, Philippines}
\date{\today}

\begin{abstract}

The self-force acting on a charged particle is sensitive to the global structure of curved spacetime and can serve as a probe of geometry beyond local curvature. We compute the static scalar self-force on a point charge in the two-parameter family of spherically symmetric wormholes introduced by Konoplya and Zhidenko, members of the broader Morris–Thorne class of traversable wormholes. Using mode-sum regularization, we analyze its dependence on the shape exponent $q$, which controls the throat geometry, and the redshift parameter $p$, which determines the redshift function and tidal strength. We find that the self-force is generally not unidirectional: it can change sign with radial distance from the throat, with up to two distinct zero crossings depending on $(p,q)$. We provide a systematic characterization of how both the direction and large-distance falloff depend on the wormhole parameters. For sufficiently large $p$, the force can decay at a slower rate than the canonical $\sim r^{-3}$ behavior typical of isolated-body spacetimes, with stronger flaring (more negative $q$) leading to more rapid decay. In the combined limit $p \to \infty$ and $q \to -\infty$, the asymptotic falloff approaches that of the static scalar self-force in the Ellis wormhole.
\end{abstract}

\maketitle

\section{Introduction}
\label{sec:intro}

Wormholes are the quintessential exotic structures of general relativity. Einstein and Rosen would first introduce them as a sheet-connecting ``bridge'' model for an elementary particle during the pre-quantum days of theoretical physics \cite{flamm1935physik}, and two decades later, Misner and Wheeler would coin their famous name \cite{misner1957gravitation}. Since then, wormholes have continued to beguile imaginations and inspire many intriguing ideas both in and out of science. In recent years, wormhole physics has enjoyed a resurgence of interest arising from many fronts. In alternative theories of gravity, wormholes have become more palatable as physical solutions, as they no longer require exotic matter for their existence and stability \cite{gravanis2007mass, richarte2007thin, eiroa2008thin, richarte2010wormholes}. As a result of this, and spurred no doubt by the incredible advances of gravitational wave astronomy, considerable effort has gone into understanding their astrophysical signatures, particularly towards assessing their potential as black hole mimickers \cite{bambi2013can, bambi2017testing, bambi2021astrophysical}. Wormhole solutions are also implicated in the black hole information paradox, where the use of replica wormholes allows one to demonstrate consistency between the Page curve of Hawking radiation and unitary \cite{almheiri2020replica}. And finally, wormholes have also taken space in condensed matter physics, where they provide fertile ground for speculative ideas in curved graphene physics \cite{gonzalez2010graphene}. Across these diverse domains, wormholes frequently emerge as geometric analogs of quantum tunneling. Notably, while defining a finite traversal time for quantum tunneling remains a notoriously subtle problem, with transmission often appearing instantaneous (\cite{flores2024instantaneous}), traversable wormholes provide a framework where the crossing time is strictly finite and causally well-behaved.    

In many of their applications, the primary appeal of wormholes lies in their nontrivial topology. The present work concerns how this nontrivial topology affects the self-force on a static particle. A restraining force needs to act on a particle that is held fixed in a curved spacetime. If this particle carries a charge that gives rise to a field, the field configuration is typically asymmetric in the local frame of the particle. In turn, this asymmetry results in a \emph{self-force} on the particle that either amplifies or diminishes the required restraining force. 

The situation is not unlike the elementary case of a point charge $q$ in the vicinity of a grounded infinite conducting plane. The presence of the conducting plane requires an external force to act on the charged particle for it to remain stationary. The self-force can be easily calculated using the method of images and is found to be attractive and proportional to $~q^2$. We attribute the force to the surface charge density that forms on the conductor, but, of course, locally, the charged particle is only aware of the field around it, which is now exerting a force on it. This is one of the simplest examples of a self-force calculation. However, simple analyses like this fail for charged point particles in curved spacetime, to the point of being unable to predict the direction of the self-force, let alone its magnitude. To know what the self-force is on a static particle in curved spacetime, one needs to calculate it explicitly. The self-force is known to depend unpredictably on the large-scale structure of the background spacetime. As a result, substantial work has gone into mapping this relationship between static self-force and spacetime structure, including wormholes. 

The self-force in curved spacetime has been the topic of intense development over the past two decades because of its significance--- it arises as a correction to the dynamics of a particle in motion \cite{vu2026self,blanco2026conservative,juraev2026electromagnetic,bini2026analytic,vaswani2026timedomainframeworkteukolskyequation,vretinaris2026numerical,blanchet2025gravitational,trestini2026constants,warburton2026gravitational,roy2025black,iglesias2025hybridization,whittall2025frequency,upton2026effective,lewis2026postadiabatic,kuchler2026self,capozziello2026analytical,leather2025inspiral}. Calculating this self-force is necessary to accurately predict the motion of particles in a given spacetime. The field generated by the particle in the background diverges at the particle's location, which poses a computational challenge. This demands a regularization procedure in which the removal of the field's singular nature does not contribute to the self-force, achieved through the Detweiler-Whiting decomposition \cite{detweiler2003self}. A regular field is then obtained after the subtraction of the appropriate singular component from the full field. Since this regular field is a solution to a homogeneous wave equation, it can be treated as an independent field interacting with the particle, and it is this interplay that produces the self-force. Comprehensive reviews of self-force and radiation reaction in general relativity are found in Refs. \cite{poisson2011motion,barack2018self}. 

Khusnutdinov and Bakhmatov initiated the earliest self-force calculation involving wormholes \cite{PhysRevD.76.124015} for an electric charge at rest in a static wormhole background. By considering singular and smooth throat profiles, it was observed that the self-force is everywhere attractive except at the throat. For an arbitrary throat profile, it was shown that the self-force is attractive and falls off as $\rho^{-3}$ far from the throat, where $\rho$ is the distance between the static particle and the throat. The attractive nature of the static self-force was argued to be an intrinsic property of the wormhole and interpreted as a consequence of its nontrivial topology. In Ref. \cite{khusnutdinov2010self}, the self-force acting on a static electric charge for the case of massive wormholes was also found to be attractive. Far from the throat, the self-force would decay as $\rho^{-3}$ (similar to \cite{PhysRevD.76.124015}) while the introduction of a mass parameter caused an asymmetry in the self-force profile. In Ref. \cite{de2012probing}, two copies of Minkowski spacetime with constant angle deficit are joined together to obtain cylindrical thin-shell wormholes. For large deficit angles, the self-force on a charged particle at rest in the wormhole was found to be repulsive or attractive, depending on the charge's location. This transition in the sign of the self-force was proposed as a determining factor in the presence of a throat. In a more recent paper by de Celis and Simeone \cite{de2020electrostatics}, asymptotic flatness was incorporated into the geometry of conical wormholes first considered in Ref. \cite{de2012probing}. When a point charge is situated between the wormhole throat and the conical throat, the magnitude of the computed repulsive force decreases with increasing value of the conical throat. The same group in \cite{de2012probing} would extend their work to the case of a thin-shell wormhole constructed from two copies of Schwarzschild geometry \cite{de2013probing}, and discover that when the throat radius is close to the value of the Schwarzschild horizon, the electrostatic self-force transitions from being repulsive to attractive as we go farther from the throat; otherwise, the self-force is attractive for all distances. For this case, the asymptotic behavior of the self-force as $\rho\rightarrow\infty$ is still $\rho^{-3}$ at the leading order. Electrostatic self-forces in wormholes thus appear to fall off universally like $\rho^{-3}$, with sign-switching across the radial direction seemingly indicative of a throat. In a more recent work \cite{aslan2022self}, Aslan and Popov analyzed the self-force for a static charge in the spacetime of a wormhole with an infinitely short throat, where it is assumed that the electromagnetic field generated by the charge is nonminimally coupled to the curvature of spacetime. The self-force diverges near the throat, a limitation of the infinitely short-throat model that the authors acknowledge, noting that smooth-throated wormhole models do not exhibit such divergences.

A similar catalogue of results has emerged for static scalar self-forces. In contrast to the electrostatic case, the scalar self-force does not only depend on the background spacetime, but also on the coupling constant $\xi$ that effectively defines the scalar field $\Psi$ with the following wave equation
\begin{equation}\label{scalar field}
\left(\square-\xi R-m^{2}\right) \Psi=-4 \pi j
\end{equation}
where $R$ is Ricci scalar, $m$ is the particle's mass, and $j$ is the scalar charge density. Bezerra and Khusnutdinov \cite{bezerra2009self} were the first to study the self-force on a scalar charge at rest in wormhole spacetimes with a specific throat profile also considered in Ref. \cite{PhysRevD.76.124015}. They showed that for a massless scalar, the self-force vanishes for all distances when $\xi=1/8$ and it diverges for $\xi=1/2$, while it becomes attractive for $\xi<1/8$ and repulsive for $\xi>1/8$. The critical dependence of the self-force on the coupling constant was further examined by Taylor \cite{taylor2013self,taylor2014propagation,taylor2017erratum}, who obtained an analytic expression for the self-force on an arbitrarily coupled static scalar charge in an ultrastatic wormhole spacetime with the throat profile $r(\rho)=\sqrt{\rho^2+a^2}$, where $r=a$ is the throat. Most intriguingly, Taylor found an infinite set of values of $\xi$ for which the self-force either vanishes or diverges. For $\xi=0$, the self-force is everywhere positive and is found to scale as $\rho^{-3}$ in the limit as $\rho\rightarrow\infty$. Additionally, in this program, cylindrically symmetric thin-shell spacetimes with an angle deficit were considered \cite{tomasini2019arbitrarily}. In a symmetric wormhole with conical regions, the self-force on a static massless scalar diverges at the radial position of the shell but becomes finite, and changes sign as we go farther away. If the angle deficit is zero, the resulting self-force has a similar profile to the conical case in the region where the particle is situated at or near the shell. The presence of a non-zero angle deficit only affects the asymptotic behavior of the self-force: the Minkowski and conical wormhole spacetimes generate repulsive and attractive forces, respectively. Meanwhile, in the massive scalar case \cite{tomasini2018self} and with minimal coupling, the self-force diverges as the particle approaches the thin-shell throat of the cylindrically-symmetric wormhole. At the same time, it becomes finite and repulsive in the vicinity of the throat. For non-minimal coupling, the self-force experiences a change in direction when $\xi=1/4$, i.e., the force tends to point towards the throat for a more positive $\xi$.  Finally, the self-force acting on a scalar charge at rest in a wormhole endowed with dilatonic charge \cite{popov2018self} was found to be everywhere attractive with respect to the throat and decay as $r^{-3}$ at leading order far from the wormhole throat. Again, the picture that is slowly emerging for scalar self-forces in asymptotically-flat wormholes is that of a universal radial falloff of $\rho^{-3}$. Similarly to the electrostatic case, sign-switching in the computed self-force has also been reported.

In this paper, we expand this body of knowledge by working with a substantially larger class of wormhole spacetimes and exploring in detail the dependence of the self-force on the physical characteristics of the wormholes. We choose to work within the class of traversable wormholes popularized by Morris and Thorne through their seminal paper \cite{morris1988wormholes1}
    \begin{equation}\label{metric}
    ds^2=-e^{2\Phi(r)}dt^2+\left(1-\dfrac{b(r)}{r}\right)^{-1}dr^2+r^2 d\Omega ^2,
    \end{equation}
which is completely determined by two freely specifiable functions $\Phi(r)$ and $b(r)$, where $d\Omega^2=d\theta^2+\sin ^2\theta d\phi^2$ is the metric on a 2-sphere. Here $\Phi(r)$ is the redshift function which determines the gravitational redshift and tidal force of the wormhole spacetime. The shape of a wormhole is fully described by another function $b(r)$, called the shape function. For the wormhole to connect distant asymptotically flat regions, the spacetime should be free from horizons, and the geometry is required to have the shape of a wormhole. These translate to the following constraints \cite{morris1988wormholes1}
\begin{itemize}
\item $\Phi(r)$ needs to be finite everywhere, i.e., $-e^{2\Phi}\neq 0$;
\item $b(r)$ must satisfy the so-called flaring-out condition described in Eq.~\eqref{flaring_out} below;
\item the geometry must contain a throat located at minimum $r=b_{0}$ such that $b(b_{0})=b_{0}$, and $1-b(r)/r\geq 0$ throughout the spacetime.
\end{itemize}
With the imposition of the flaring-out condition, the matter and energy content making up the Morris-Thorne wormhole solution of GR fundamentally violates the null energy condition (and all energy conditions \cite{lobo2017wormholes}), which, from a classical point of view, is problematic. However, some quantum effects lead to small but experimentally measurable violations of the null energy condition \cite{krasnikov2000traversable, garattini2005self,garattini2009self,kuhfittig2018traversable}. Other possible sources of exotic matter include scalar fields and cosmic phantom energy \cite{armendariz2002class,sushkov2005wormholes,lobo2005phantom}. It remains an interesting problem whether or not it is possible to obtain enough exotic matter to support a macroscopic traversable wormhole since all known violations are extremely small, while the presence of exotic matter also poses a question on the wormhole's stability \cite{lobo2010closed}. Efforts to minimize the amount of required exotic matter have become a common preoccupation of wormhole research \cite{visser1997geometric,debenedictis2001general,rahaman2006theoretical,forghani2018fate}. Other works also aimed to fully circumvent this problem by exploring wormhole solutions in alternative theories of gravity where exotic matter need not be present (see Ref. \cite{harko2013modified} for a summary).

Among traversable wormholes, we focus on the two-parameter family of wormhole solutions introduced by Konoplya and Zhidenko \cite{konoplya2010passage}
	\begin{equation}\label{class}
		b(r)=b_0^{1-q}r^q ,\qquad \Phi(r)=\frac{\sigma b_0^p}{r^p} 
    \end{equation}
where $q<1$ and $p>0$, and $\sigma = \pm 1$. (The original parametrization only has $\sigma =+1$, but as we explain below, it is also interesting to include the $\sigma = -1$ case.) Ultrastatic wormholes are those for which $\Phi=0 $ (no redshift and zero tidal force), and are represented by $p=0$ in this parametrization. An ultrastatic wormhole with $q=-1$ corresponds to the well-known Ellis wormhole \cite{ellis1973ether}. The traversable wormholes described in Eqs.~\eqref{metric} and \eqref{class} reduce to two copies of Minkowski spacetime in the limit
\begin{equation}\label{minkowski limit}
\lim\limits_{q\rightarrow-\infty}b(r)=0,\quad \lim\limits_{p\rightarrow\infty}\Phi(r)=0.
\end{equation}
We shall refer to this as the KZ class of wormholes or just KZ wormholes. The exact form of the shape function in Eq.~\eqref{class} was later used in Refs. \citep{cataldo2015morris,godani2019traversable} in the study of traversable wormholes and energy conditions while nearly all of the shape functions considered in the literature dealing all aspects of wormhole physics can be modelled in the form of the KZ wormholes via the shape exponent $q=1/2$ (see Refs. \citep{morris1988wormholes1,konoplya2016wormholes,konoplya2018tell}) and $q=-1$ (see Refs. \cite{PhysRevD.76.124015,bezerra2009self,taylor2013self,kim2013flare,roy2020revisiting,azad2020transmission,churilova2020arbitrarily}).

Calculating the self-force in this family of wormholes does not appear to be possible to do fully analytically. Fortunately, there exist well-developed practical methods
for numerically calculating the self-force \cite{casals2012regularization,Vega_2011,wardell2014self}. Among these, the mode-sum regularization procedure \cite{barack2000mode} is the most appropriate method for calculating static self-forces. The most essential part of the recipe for all these practical methods is the removal of an undesirable singular part from the full field (as per the Detweiler-Whiting decomposition). In mode-sum regularization, spherical-harmonic modes of the full field are numerically calculated, and the modes of the singular field are analytically determined. Regularization is then a matter of subtracting analytic singular-field modes from the numerical full field modes to leave the pieces that contribute to the self-force \cite{burko2000self,barack2003gravitational,haas2006mode,heffernan2012high}. This method allows us to fully explore the behavior of the static scalar self-force in KZ wormholes. 

Several new interesting results arise from this exploration. First, the scalar self-force can also exhibit sign-switching behavior similar to previous results for electrostatic and scalar self-forces. But whether or not this sign-switching occurs does not critically depend on the existence of a throat, but rather on the specific parameters of the wormhole. We establish, in particular, that no ultrastatic wormhole can exhibit this sign-switching feature. Second, even though all of the KZ wormholes are asymptotically flat, the large-distance behavior of the self-force is not the typical cubic falloff, but also depends on the KZ parameters. In general, we find that both the redshift parameter $p$ and shape exponent $|q|$ increase the magnitude of the self-force. 
For large $p$, the self-force is found to decay faster for a wormhole with a shape that exhibits a greater degree of flaring out. In the limit as $p\rightarrow\infty$ and $q\rightarrow-\infty$, the falloff asymptotes to the Ellis wormhole result $\rho^{-3}$.

The paper is organized as follows. In Sec.~\ref{sec:wh}, we provide a short review of static trajectories and gravitational redshift in wormholes and a derivation of the general expression for the throat profile. In Sec.~\ref{sec:fe}, derivation of the field equation describing the static scalar charge in the wormhole spacetime is provided. Sec.~\ref{sec:msrp} is written for a discussion on the regularization procedure and mode-sum formalism used in computing the self-force. Finally, we present in Sec.~\ref{sec:results} numerical results and compare the ultrastatic case with that obtained in Ref. \cite{taylor2013self}.

Throughout this paper, we adopt geometric units $G=c=1$ and a mostly plus signature.

\section{Traversable wormholes}
\label{sec:wh}

For completeness, we describe some of the properties of the class of wormhole spacetimes studied here. We work with parametrized Morris-Thorne wormholes introduced by Konoplya and Zhidenko, given in Eq.~\eqref{class}. In what follows, the characteristic flaring-out condition, gravitational redshift, and static trajectories in these wormholes are discussed in terms of the KZ parameters $(p,q)$.

\subsection{General throat profile}
It is usual to express the metric in terms of the proper distance $\rho$ defined by 
\begin{equation}\label{properdistance}
\rho ( r ) = \pm \int _ { b _ { 0 } } ^ { r } \frac { d\,\overline{r}} { \sqrt { 1 - b(\overline{r}) / \overline{r}}},
\end{equation}
with which the metric in Eq. \eqref{metric} takes the form
\begin{equation}\label{metric1}
ds^2=-e^{2\Phi}dt^2+d\rho^2+r^2(d\theta^2+\sin ^2\theta d\phi^2),
\end{equation}
where $\Phi=\Phi(\rho)$ and $r=r(\rho)$. There are two infinities $\rho=\pm \infty$ at $r=\infty$, which correspond to two asymptotically flat regions of spacetime. In this coordinate, the throat is now located at $\rho=0$.

\begin{figure*}[htp]
  \centering
  \subfigure[\label{fig:embedding}]{\includegraphics[scale=0.32]{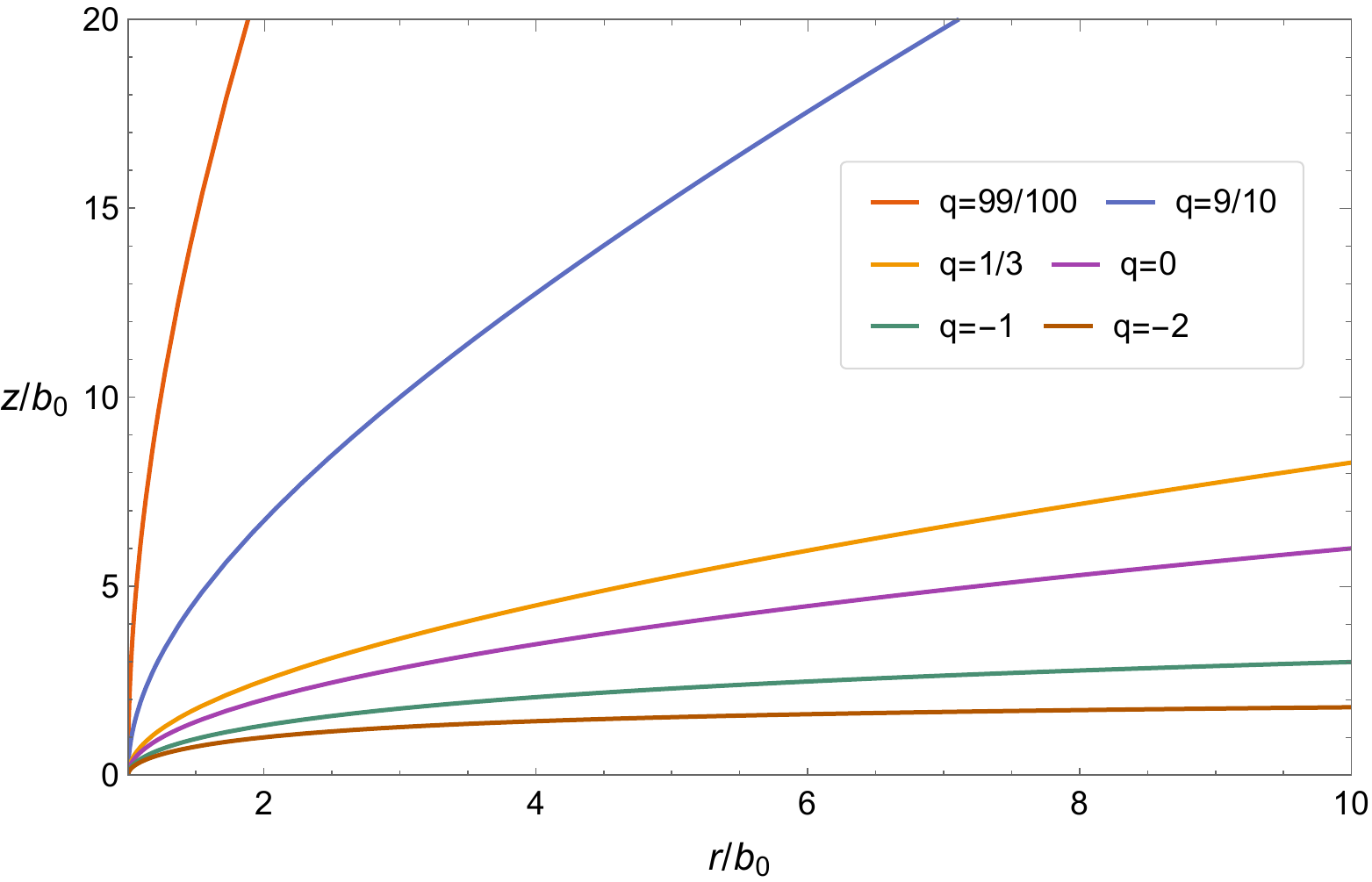}}\qquad
  \subfigure[\label{fig:redshift}]{\includegraphics[scale=0.311]{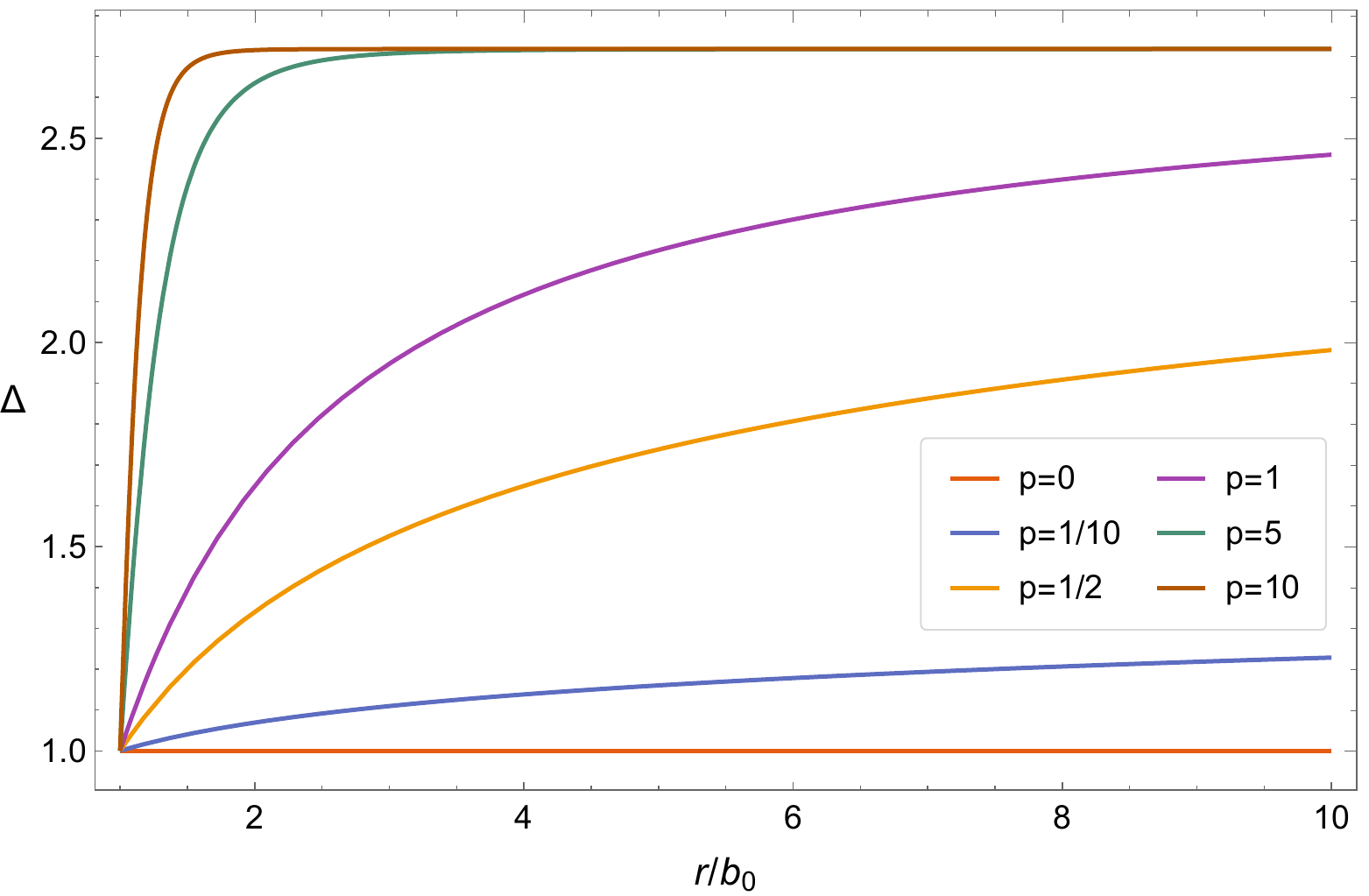}}\quad
  \caption{(a) Embedding diagram showing the characteristic flaring-out of the wormhole geometry for various shape exponents $q$, wormhole flares out more with smaller value of $q$; (b) the degree of blueshifting of light emitted at the throat $b_{0}$ increases with increasing $p$, while static observer at the throat and observer situated anywhere in an ultrastatic wormhole detects no shift in the frequency of light.}
  \label{fig:parameters}
\end{figure*}

For a KZ wormhole of arbitrary shape exponent $q$, the transformation in Eq.~\eqref{properdistance} can be expressed in terms of a hypergeometric function (see Appendix \ref{appendix a} for derivation)
\begin{equation}\label{coord-trans}
    \rho ( r ) = \frac { 2 b _ { 0 } } { 1 - q } \zeta^{1/2}\, _ { 2 } F _ { 1 } \left( \frac { 1 } { 2 } , \frac { q - 2 } { q - 1 } ;\frac { 3 } { 2 } ; \zeta \right),
\end{equation}
where $ \zeta=1-\left(b_{0}/r \right) ^ { 1 - q }$. The expression of proper radial distance in Eq.~\eqref{coord-trans} is in agreement with that obtained in Ref. \cite{taylor2014propagation}, albeit in form. In our work, we shall frequently make use of a nondimensionalized version of this relation in terms of $y = \rho/b_0$ and $\tilde{x}=r/b_0$:
\begin{equation}\label{coord-trans-nondim}
	y(\tilde{x}) = \dfrac{2}{1-q}\zeta^{1/2}\, _ { 2 } F _ { 1 } \left( \frac { 1 } { 2 } , \frac { q - 2 } { q - 1 } ;\frac { 3 } { 2 } ; \zeta \right),
\end{equation}
where  $ \zeta=1-\left(1/\tilde{x} \right) ^ { 1 - q }$. This makes explicit the fact that the relation between areal radius $r$ and proper distance from the throat $\rho$ is solely determined by the shape parameter $q$, and that $b_0$ only serves as an arbitrary length scale.

For later use in the wave equation, it is often convenient to work with the inverse of Eqs.~\eqref{coord-trans} or \eqref{coord-trans-nondim}, $r(\rho)$ (or $\tilde{x}(y)$). But an exact analytic expression for $r(\rho)$ cannot generally be obtained.
This inversion can be carried out explicitly for shape exponents $q=-1$ and $q=1/3$, but we are unable to find any more. The shape exponent $q=-1$ corresponds to the most commonly used throat profile in the literature, which is given by
\begin{equation}\label{tp}
    r(\rho)=\sqrt[]{\rho^2+b_{0}^2},\,\, \textrm{or} \,\, \tilde{x}(y) = \sqrt[]{y^2+1}.
\end{equation}
Meanwhile, the $q=1/3$ profile described by
\begin{align}\label{tp1}
   \tilde{x}(y)=\left[y^2-5+\frac{3\cdot 2^{1/3}}{\xi}\left(1-2y^2+\frac{\xi^2}{2^{2/3}}\right)\right]^{1/2}
\end{align}
where
\begin{equation}
    \xi=\left(\sqrt{y^2(4+y^2)^3}+2+10y^2-y^4\right)^{1/3},
\end{equation}
does not seem to have been studied before in the literature.

For generic KZ wormholes (i.e., generic pairs $(p,q)$), we obtain the inverse of Eq.~\eqref{coord-trans}, $r(\rho)$, numerically. For these cases, our metric functions in terms of $\rho$, namely  $r(\rho)$ and $\Phi(r(\rho))=\sigma b_0^p/r(\rho)^p$ that appear in Eq.~\eqref{metric1}, will be numerical. The use of numerical metrics as a background for self-force calculations appears to have never been done before, likely because of the delicate nature of self-force regularization, which may be easily spoiled by inaccuracies in numerical metrics. We demonstrate in Sec.~\ref{subsec:num} that this is not a problem in our calculations. 

The wormhole can be visualized through an embedding function $z(r)$, which can be obtained from
\begin{equation}\label{embedding function}
\frac{dz}{dr}=\pm\left(\left(\frac{r}{b_0}\right)^{1-q}-1\right)^{-1/2}
\end{equation}
with a manifest form calculated in Ref. \cite{taylor2014propagation}.
Inspecting Fig. \ref{fig:embedding}, a wormhole with a throat profile given in Eq.~\eqref{tp1} has a longer spatial extension of the throat regime but flares out less than that of a wormhole with throat profile in Eq.~\eqref{tp}. 

Using embedding diagrams, we can extract useful information for choosing the shape function $b(r)$. To be a wormhole, the geometry has a minimum radius, $r = b ( r ) = b _ { 0 }$, denoted as the throat, at which the embedded surface is vertical, i.e., $d z / d r \rightarrow \infty$. The throat connects two regions of spacetime which we may consider as asymptotically flat far from the throat, $d z / d r \rightarrow 0$ as $r\rightarrow \infty$. Near the throat, the embedding function in Eq.~\eqref{embedding function} flares out. Alternatively, the flaring-out condition is written in terms of the inverse of the embedding function 
\begin{equation}\label{flaring_out}
\frac{d^2r}{dz^2}=\frac{b(r)-rb'(r)}{2b^2(r)}>0,
\end{equation}
which must be concave up at or near the throat. The flaring-out condition in Eq.~\eqref{flaring_out} gives a constraint on the form of the shape function $b(r)$ and its allowed parametrization. More importantly, this condition aids in the understanding of how the exotic matter content of traversable wormholes violates the energy conditions \cite{lobo2017wormholes}.

\subsection{Gravitational redshift and static trajectories}
The ratio between the measured frequency $\omega_{2}$ (at $r_2$) and transmitted frequency $\omega_{1}$ (at $r_1$) of signals involving two static observers in the vicinity of the wormhole is given by 
    \begin{equation}\label{redshift}
        \Delta=\frac{\omega_{2}}{\omega_{1}}=\sqrt{\frac{g_{tt}|_{r_1}}{g_{tt}|_{r_2}}}= \exp\left[\sigma b_0^p\left(\dfrac{1}{r_1^p}-\dfrac{1}{r_2^p}\right)\right]
    \end{equation}
hence $p$ is referred to as the redshift parameter. Therefore, as a photon moves away from the wormhole throat ($r_2 > r_1$), if $\sigma = +1$, we have $\Delta > 1$, and we see the measured frequency $\omega_2$ blueshifted compared to the transmission frequency $\omega_1$. We see then that the case $\sigma=+1$ represents a wormhole that behaves like a rather peculiar gravitational ``center", one in which a photon ``falls down" a gravitational well and becomes more energetic as it moves away from the center. We need $\sigma=-1$ to recover wormholes that act on photons in the more usual way.  

For a small outward displacement $\delta r$, say from $r_1 = r$ to $r_2=r+\delta r$, the frequency shift experienced by the photon is 
    \begin{align}
        \Delta &= 1 - 2\Phi'(r) \delta r + O((\delta r)^2) \\ &= 1+  \sigma p\, b_0^p\,r^{-(p+1)} \delta r + O((\delta r)^2).
    \end{align}
Again, this tells us that $\sigma = +1$ gives a blueshift, but also that $p$ (through $\Phi'(r)$) controls the degree of blueshifting. For $\sigma = +1$, larger $p$s give stronger blueshifts, and for $\sigma = -1$, larger $p$s give stronger redshifts. It is for this reason that we call $p$ the redshift parameter. In Fig.~\ref{fig:redshift}, we plot the blueshift experienced by a photon emitted at the throat ($r_1=b_0$) as it gets to $r_2 = r$. As $r\rightarrow \infty$, the maximum blueshift (for all $p$) is $\Delta = e$. Therefore, in any KZ wormhole with $\sigma = +1$, outward photons all eventually get blueshifted by a factor of $e$. Similarly, for any KZ wormhole with $\sigma = -1$, outward photons all get redshifted by a factor of $1/e$ as they reach infinity. The figure also shows how rapidly the frequency varies when $p$ is altered. 

The role of $\sigma$ can be further understood by considering static particles in KZ wormholes. A static particle will have a four-velocity
$u ^ { \alpha } = d x ^ { \alpha } / d \tau = \left( e ^ { - \Phi ( r ) } , 0,0,0 \right)$ and four-acceleration with components $a^{t}=0$ and 
\begin{align}\label{4-acc}
a ^ { r } =& \,\Phi ^ { \prime } \left( 1 - \frac { b(r) } { r } \right)\nonumber\\
=&-\sigma p b_0^p\,r^{-(p+1)} \left[1-\left(\frac{b_0}{r}\right)^{1-q}\right]
\end{align}
for arbitrary shape exponent $q$ and redshift parameter $p$, where prime denotes differentiation with respect to $r$. 
Therefore, any static particle at the throat  $r = b ( r ) = b _ { 0 }$ for any $\Phi(r)$ is moving along a geodesic. Meanwhile, if the redshift function is constant, i.e., $\Phi'(r)=0$, then the motion of static particles fixed at any $r$ is also geodesic. In general, however, static particles are accelerating. For the case of Schwarzschild, for which $b(r) = 2M$ and $\Phi(r) = (1/2)\log(1-2M/r)$, $a^r = M/r^2 > 0$. The acceleration is outward because the geodesic trajectories of free-falling particles go inward. The sign of the acceleration can be interpreted to be the direction in which an external force acting on the particle must be pointing in order to keep the particle at a fixed position. 

Because $a^r > 0$, the Schwarzschild geometry can be said to be an attractive gravitational center. 
Similarly, if $a^{r}>0$ for a static particle in a wormhole, then the latter can be said to be attractive. On the other hand, the wormhole is repulsive if $a^{r}<0$. From Eq.~\ref{4-acc}, when $\sigma=+1$, the wormhole is repulsive. For $\sigma=-1$, the wormhole is attractive.

\section{Field equations}
\label{sec:fe}

A scalar field $\Psi$ sourced by a scalar charge $Q$ can be considered to be a small perturbation on the static wormhole background. This is described by the wave equation
    \begin{equation}\label{wave equation}
   \frac{1}{\sqrt[]{-g}}\partial_{\mu}\left(\sqrt[]{-g}g^{\mu\nu}\partial_{\nu}\Psi\right)=-4\pi j
    \end{equation}
with metric determinant $g=-e^{2\Phi}r^4\sin^2\theta$.
For the charge at rest at point $x_s$ according to observers moving along integral curves of the Killing vector $\partial/\partial t$, the source density amounts to
\begin{align}
j(x)=&\,Q\int \delta^{(4)}(x-x_s(\tau))\frac{d\tau}{\sqrt[]{-g}}\\
=&\frac{Q}{r^2 \sin\theta}\delta^{(3)}(x-x_s).
\end{align}

\subsection{Mode-decomposition}

In order to integrate the field equation, we implement a multipole decomposition of the scalar potential
\begin{equation}
\Psi(\rho,\theta,\phi)=\sum_{lm}\psi_{lm}(\rho)Y_{lm}(\theta,\phi)
\end{equation}
and in a similar way, the source is decomposed into
\begin{equation}
j(x)=\frac{Q\,\delta(\rho-\rho_s)}{r^2}\sum_{lm}Y^*_{lm}(\theta_s,\phi_s)Y_{lm}(\theta,\phi)
\end{equation}
where $r=r(\rho)$, in which $(\rho_s,\theta_s,\phi_s)$ represent the spherical coordinates of the position $x_s$ of the source.

Without loss of generality, we may place the particle along the polar axis $(\theta_s=0)$ and exploit the property of spherical-harmonic functions
    \begin{equation}
    Y_{lm}(0,\phi)=\sqrt[]{\frac{2l+1}{4\pi}}\delta_{m,0}.
    \end{equation}
After which, substitution within the field equation produces
\begin{align}\label{eqn in r}
 r^2\psi''+&\left(\Phi'+\frac{2r'}{r} \right)r^2\psi'-l(l+1)\psi\nonumber\\
 &=-4\pi Q\,\sqrt[]{\frac{2l+1}{4\pi}}\delta(\rho-\rho_s)
\end{align}
where $\psi=\psi_{l0}(\rho)$ and with the prime here denoting differentiation on $\rho$. The modes with spherical harmonic index $m\neq 0$ necessarily vanish. We then express Eq.~\eqref{eqn in r}  in terms of dimensionless parameters $y=\rho/b_0$ and $\tilde{x}=r/b_0$,
which takes the final form
\begin{align}\label{wave eqn in y}
\tilde{x}^{2}\psi''(y)+&\left(2\tilde{x}-\frac{\sigma\,p}{\tilde{x}^{p-1}}\right)\,\tilde{x}'\psi'(y)-l(l+1)\psi(y)\nonumber\\
&=-4\pi \frac{Q}{b_0}\,\sqrt[]{\frac{2l+1}{4\pi}}\delta(y-y_s)
\end{align}
with the prime denoting differentiation on $y$ (see Appendix \ref{appendix b} for the derivation). This shows that $\psi \sim Q/b_0$, i.e., the physical/ dimensionful quantities $Q$ and $b_0$ are only able to rescale $\psi$, which implies that for our numerical calculations, there is no loss of generality in setting $Q=1$ and $b_0=1$. In this final form, we note that $\tilde{x}$ is a function of $y$. For KZ wormholes, an explicit form for $\tilde{x}(y)$ is only available for $q=-1$ and $q=1/3$. For generic KZ wormholes, $\tilde{x}(y)$ is determined numerically by inverting Eq.~\eqref{coord-trans-nondim}.

\subsection{Boundary conditions}
The general solution to the scalar wave equation in Eq.~\eqref{wave eqn in y} may be written as a sum of two linearly independent solutions 
\begin{equation}\label{ansatz1}
\psi=c_{1}\,\psi_{<}\,\Theta(y_s-y)+c_{2}\,\psi_{>}\,\Theta(y-y_s)
\end{equation}
of the homogeneous differential equation
\begin{align}\label{homo-eq1}
    \tilde{x}^{2}\psi''(y)+\left(2\tilde{x}-\frac{\sigma\,p}{\tilde{x}^{p-1}}\right)\,\tilde{x}'\psi'(y)-l(l+1)\psi(y)=0
\end{align}
where
\begin{equation}
\psi= \begin{cases}c_{1} \psi_{<}, & \text {if } y<y_{s} \\ c_{2} \psi_{>}, & \text {if } y>y_{s}\end{cases}
\end{equation}
and $\Theta$ is the Heaviside step function. We can obtain the jump condition at the location of the charge $y=y_s$ by combining Eqs.~\eqref{ansatz1} and \eqref{wave eqn in y}. Matching terms with the delta function, we arrive at the following junction condition
\begin{equation}
\left[c_2\psi_>'-c_1\psi_<'\right]_{y=y_s}=-\frac{Q}{b_{0}\tilde{x}_{s}^2}\sqrt[]{4\pi(2l+1)}
\end{equation}
where $\tilde{x}_{s}=\tilde{x}(y_{s})$. The solutions must also be continuous at the boundary
\begin{equation}\label{continuity1}
c_1\psi_<(y_s)=c_2\psi_>(y_s)
\end{equation}
where the amplitude $c_1$ can be obtained from the jump condition as
\begin{equation}\label{jump-cond1}
c_1=-\frac{Q \, \sqrt[]{4\pi(2l+1)}}{b_{0}\tilde{x}_{s}^2}\left(\frac{\psi_>}{\psi_<\psi_>'-\psi_<'\psi_>}\right),
\end{equation}
while $c_2$ is solved from the continuity condition as
\begin{equation}
    c_2=-\frac{Q \, \sqrt[]{4\pi(2l+1)}}{b_{0}\tilde{x}_{s}^2}\left(\frac{\psi_<}{\psi_<\psi_>'-\psi_<'\psi_>}\right).
\end{equation}
The functions $\psi_<$ and $\psi_>$ are all evaluated at the particle location $y=y_s$.

The appropriate boundary conditions on the scalar field at $y=\pm \infty$ require
\begin{align}
    &\psi_<(y)\xrightarrow{} 0 , \hspace{1cm} \text{as} \hspace{0.2cm} y\xrightarrow{}-\infty\\
    &\psi_>(y)\xrightarrow{} 0, \hspace{1cm} \text{as} \hspace{0.2cm} y\xrightarrow{}+\infty
\end{align}
i.e., the fields in the two asymptotically flat regions must vanish.
Here, the necessary boundary value of the field at infinity can be obtained by performing the method of Frobenius, which assumes a power series solution of the form
\begin{equation}\label{asymp}
    \psi(y)=\dfrac{1}{y^k}\sum_{i=0}^{\infty}\dfrac{a_i}{y^i}
\end{equation}
around $y=\infty$.  The solution is built on the expansion coefficients $a_i$, which are functions of the $l$-modes. For $ q=-1 $, we obtain a recurrence relation for $ a_i $ in terms of the redshift parameter $ p=n/m $ (see Appendix \ref{appendix c} for the derivation)

\begin{align}
a_{i}=&\left(i (i+2 l m+m)\right)^{-1}\left[(i+(l-1) m)^2 a_{i-2 m}\right.\nonumber\\
&-n (i+l m+m-n) a_{i-n}\nonumber\\
&\left.+n (i+(l-1) m-n) a_{i-2 m-n}\right]
\end{align}
where $n,m\in\mathbb{N}$.
The asymptotic solution, truncated to some order, and its derivative provide the necessary boundary value of the scalar field at an approximate infinity acting as initial data in the numerical integration of Eq. \eqref{wave eqn in y}.

\begin{figure*}[htp]
  \centering
  \subfigure[\label{fig:falloff ultrastatic}]{\includegraphics[scale=0.331]{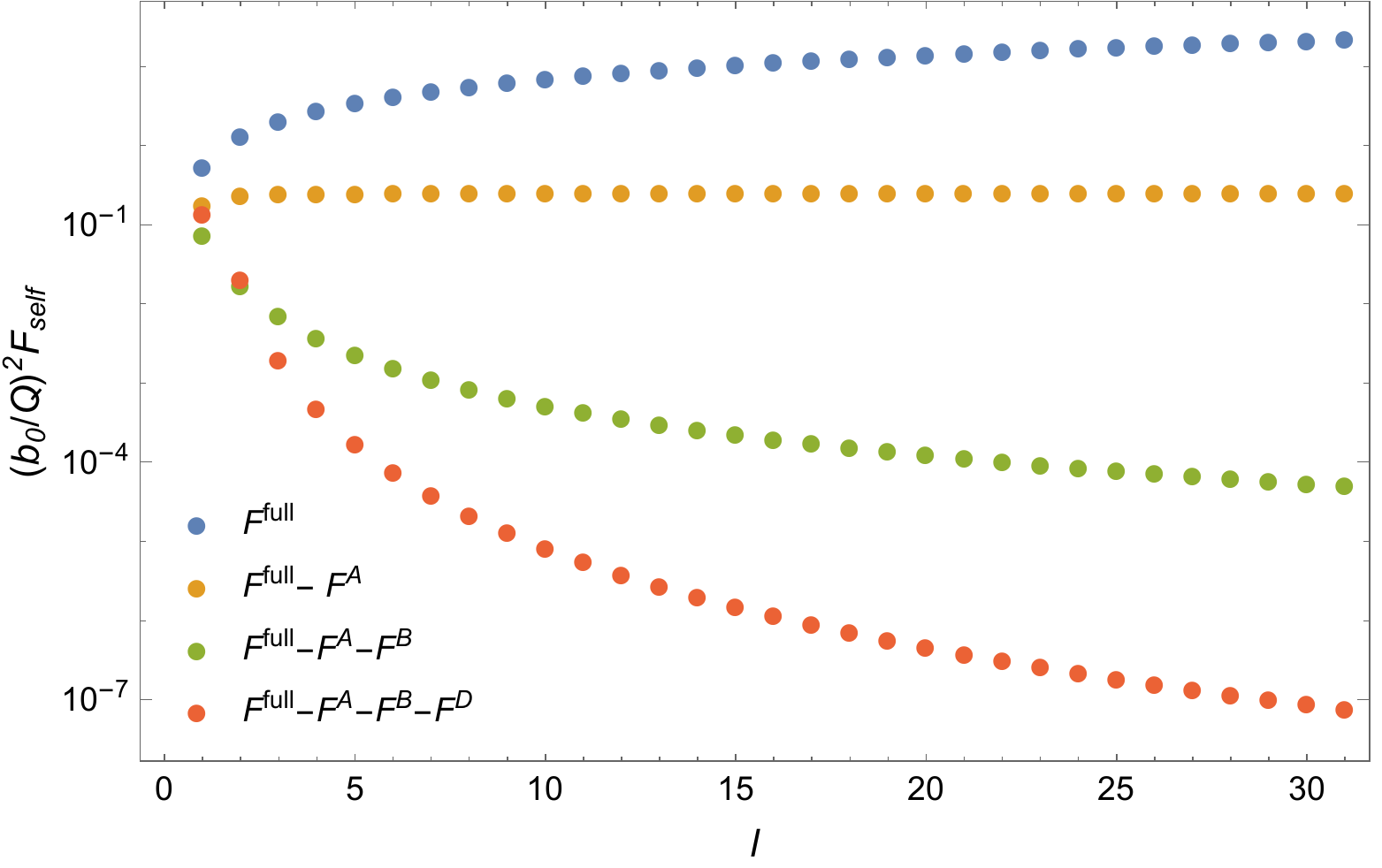}}\quad
  \subfigure[\label{fig:sf ultrastatic}]{\includegraphics[scale=0.331]{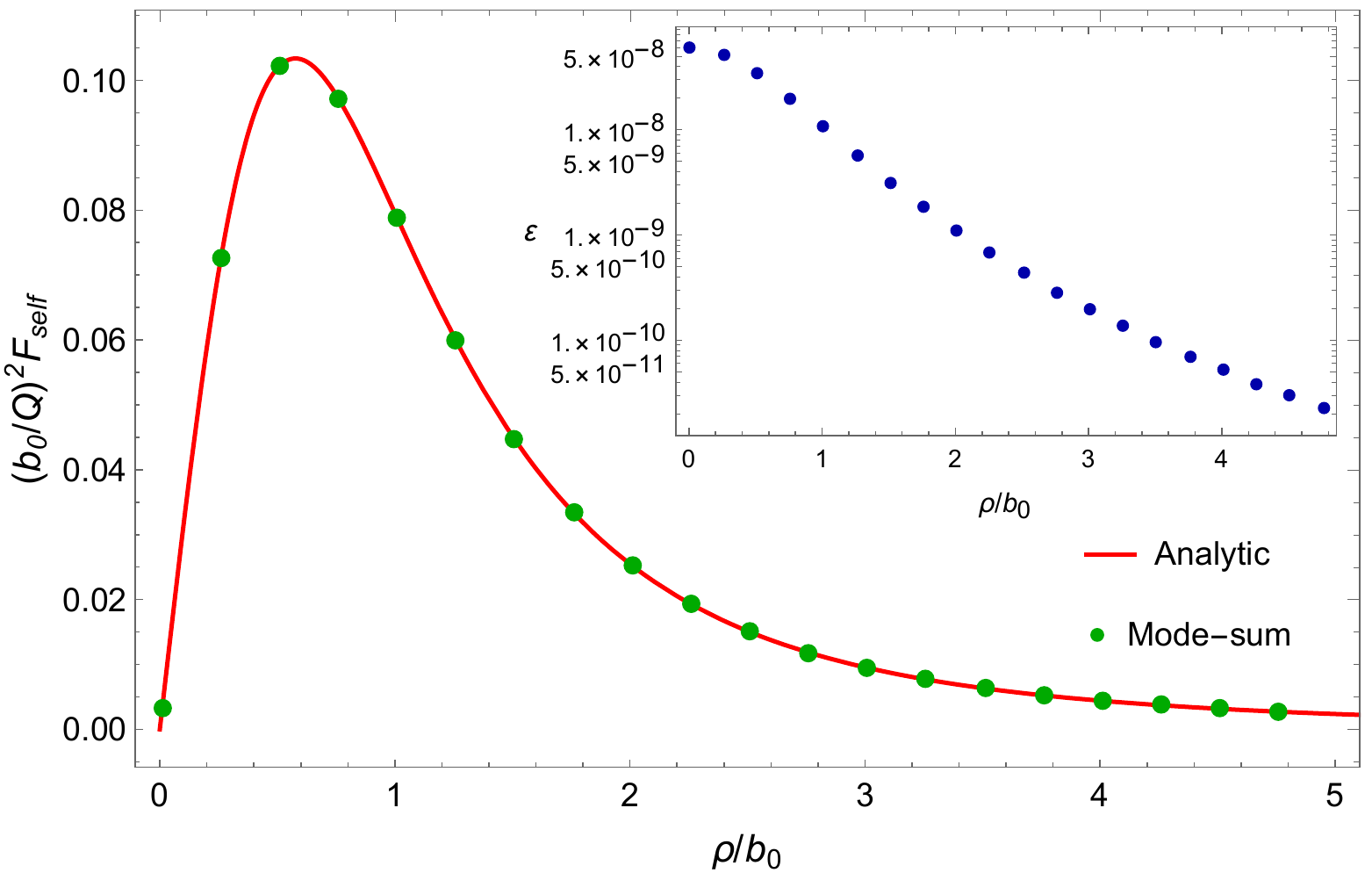}}
  \caption{(a) Falloff behavior of the $l$-mode components of the self-force on the scalar charge located at $\rho/b_0=1$ along with results of the subtraction of $A$, $B$ and $D$ regularization parameters; (b) comparison of the self-force obtained using Taylor's exact expression, and by the mode-sum approach for the ultrastatic wormhole, (Inset) log plot of the relative error $\varepsilon$ between the self-force analytically obtained by Taylor and numerically obtained using mode-sum approach (with maximum mode number $l_{\mathrm{max}}=30$) as a function of the particle location.}
  \label{fig:ultrastatic case}
\end{figure*}

\section{Mode-sum regularization}
\label{sec:msrp}
The self-force is extracted from the retarded field, which is the gradient of the full potential $\Psi^{\mathrm{full}}$, and it diverges at the position of the particle. This requires regularization, which can be carried out by implementing Detweiler-Whiting decomposition \cite{detweiler2003self}, which prescribes a unique decomposition of the divergent full potential into its regular $\Psi^{\mathrm{R}}$ and singular $\Psi^{\mathrm{S}}$ pieces
\begin{equation}
    \Psi^{\mathrm{full}}=\Psi^{\mathrm{R}}+\Psi^{\mathrm{S}}.
\end{equation}
This amounts to computing the Detweiler-Whiting singular field, which, upon subtraction from the full field, returns a regular piece, i.e.,
\begin{equation}\label{DW decomp}
    F _ { \alpha }^{\mathrm{self}} = Q \nabla _ { \alpha } \Psi ^ { \mathrm { R } } = Q \left( \nabla _ { \alpha } \Psi^{\mathrm{full}} - \nabla ^ { \alpha } \Psi ^ { \mathrm { S } } \right).
\end{equation}
Both $\Psi^{\mathrm{full}}$ and $\Psi ^ { \mathrm { S } }$ diverge at the position of the particle with charge $Q$--- any kind of subtraction will be impractical since it still reproduces the singular nature of the self-force. This difficulty is resolved by applying mode-decomposition and by expressing these singular quantities in terms of finite ones. The mode-sum formula for the $\alpha$-component of the self-force is given by \cite{poisson2011motion}
\begin{equation}\begin{aligned}
F_{\alpha}^{\text {self }}=& Q \sum_{l=0}^{N}\Big[\left(\nabla_{\alpha} \Psi^{\text {full }}\right)_{l}-\left(l+\frac{1}{2}\right) A_{\alpha}-B_{\alpha}-\frac{C_{\alpha}}{l+\frac{1}{2}}\\
&-\frac{D_{\alpha}}{\left(l-\frac{1}{2}\right)\left(l+\frac{3}{2}\right)}-O\left(\ell^{-4}\right)\Big]+\text {remainder }
\end{aligned}\end{equation}
requiring certain analytic functions called the regularization parameters represented by $A_\alpha$, $B_\alpha$, $C_\alpha$, $D_\alpha$. It is necessary to truncate the infinite sum up to a maximum mode number $N$ since the modes are numerically obtained, whereas the remainder terms contain all excess pieces of the $N$-summed modes. The regularization parameters for a static scalar charge in any static, spherically symmetric spacetime were computed by Casals, Poisson, and Vega \cite{casals2012regularization}. In the spacetime of a traversable wormhole described in terms of the metric functions in Eq.~\eqref{metric1}, the regularization parameters are written as
\begin{align}
   A_r=&-\frac{1}{r^2 \sqrt{1-\tilde{x}^{q-1}}},\,\, B_r=\frac{\sigma p\,\tilde{x}^{-p}-1}{2 r^2},\,\, C_r=\,0\\
    D_r=&\,\frac{1}{16} r^{-3 (p+1)} \tilde{x}^{q} \Big[\sigma p b_{0}^{p}r\tilde{x}^{-q} \Big(2\left(p^{2}-1\right) r^{2 p}\nonumber\\
    &+ \sigma p b_{0}^{p}\left(3 r^{p}- \sigma p b_{0}^{p}\right)\Big)+ b_{0}\Big((-2+q) q r^{3 p}\nonumber\\
    &+ \sigma p b_{0}^{p}\left(\left(2-2 p^{2}+3 p(-1+q)-(-2+q) q\right) \right.\nonumber\\
    &\left.\times\,r^{2 p}+ \sigma p b_{0}^{p}\left(-3 r^{p}+ \sigma p b_{0}^{p}\right)\right)\Big)\Big]
\end{align}
denoting $\tilde{x}=r/b_{0}$  where $r=r(\rho)$.

In a static setting, the self-force only has a conservative piece since the time component vanishes, and only the spatial components survive. In a spherically symmetric spacetime, the angular components also vanish, and so we are left with a radial component given by \cite{casals2012regularization}
\begin{equation}
      F_{y}^{\mathrm{self}}=\dfrac{Q^2}{b_0^2} f_{s}\frac{d\bar{\Psi}^{\mathrm{R}}}{dy}\bigg\rvert_{y=y_s},
\end{equation}
where $y_s$ is the radial position of the particle and $f_{s} := 1-b(\tilde{x}(y_s))/\tilde{x}(y_s)$. 

\section{Self-force in KZ wormholes}
\label{sec:results}

In this section, we present our results for the static scalar self-force in various KZ wormholes. The quantity we wish to compute and characterize is 
\begin{equation}
	F_y^{\text{self}} = \dfrac{Q^2}{b_0^2} f(y; \sigma, p,q)
\end{equation}
where $y=\rho/b_0$, and $\sigma, p, q$ are the wormhole parameters. 

We shall first focus on two classes of KZ wormholes ($q = -1, 1/3$) for which the metric functions appearing in Eq.~\eqref{metric1} are exact. Within each class, we explore the dependence of the self-force on the redshift parameter $p$. Belonging to the $q=-1$ class is the Ellis wormhole, an ultrastatic wormhole in which an exact expression for the static self-force was obtained in \cite{taylor2013self}. There is nothing particularly special about the $q=-1$ and $q=1/3$ classes of KZ wormholes, except that they happen to have exact expressions for the metric functions when expressed in terms of proper distance from the throat. However, they already possess the range of new behaviors of the static self-force that we wish to showcase.  

After these two cases, we move to generic KZ wormholes and study the dependence of the static self-force on general parameters $(p,q)$.  

\subsection{$(q=-1)$ wormholes}\label{SecA}
This is the most commonly studied class of wormholes. They have a throat profile given in Eq.~\eqref{tp}: 
\begin{equation*}
r(\rho)=\sqrt[]{\rho^2+b_{0}^2}.
\end{equation*}

\subsubsection{Ellis wormhole $(p=0, q=-1)$}
The ultrastatic wormhole $(\Phi=0)$ belonging to this class is the well-known Ellis wormhole. The self-force on a static scalar charge $Q$ has an exact form that was obtained by Taylor \cite{taylor2013self}. For the case of a minimum coupling, the self-force is given by
\begin{equation}\label{taylor}
    F_{\rho}^{\mathrm{Ellis}}=\frac{Q^{2} b_{0}\,\rho}{\pi\left(\rho^{2}+b_{0}^{2}\right)^{2}} = \dfrac{Q^2}{\pi b_0^2}\dfrac{y}{(1+y^2)^2},
\end{equation}
hence 
\begin{equation}
f(y; p = 0,q = -1) = \dfrac{y}{\pi(1+y^2)^2}. 
\end{equation}
The static scalar self-force in the Ellis wormhole is therefore always repulsive. It vanishes at the throat ($y=0$) and scales as $F_{\rho}^{\mathrm{Ellis}} \sim y^{-3}$, as $y\rightarrow \infty$. As is always the case in the scalar sector, the self-force is proportional to the square of the charge of the particle. 

Our numerical mode-sum code is able to successfully reproduce Taylor's exact result. This is shown in Fig.~\ref{fig:ultrastatic case}. The standard $l$-mode falloff behavior of the self-force is shown in Fig.~\ref{fig:falloff ultrastatic} with varying degrees of regularization. The singular nature of the full field is evident through the blue points, i.e., the $l$-mode components of the self-force grow linearly with increasing $l$-modes. As expected, the sum diverges. After carrying out the first regularization, which is linear in $l$, the mode components become constant, but the sum still diverges. The value of the components decreases just after the second regularization through $B$, which removes the divergent constant terms from the field, resulting in a falloff scaling as $l^{-2}$. The sum of the $l$-mode components rapidly converges when the third regularization proportional to $l^{-2}$ is applied, leaving a falloff rate of $l^{-4}$.

\begin{figure*}[htp]
  \centering
  \subfigure[\,\,$0 < p\leq 1/2\,\, (q=-1,\sigma = +1)$\label{fig:sf case1}]{\includegraphics[scale=0.33]{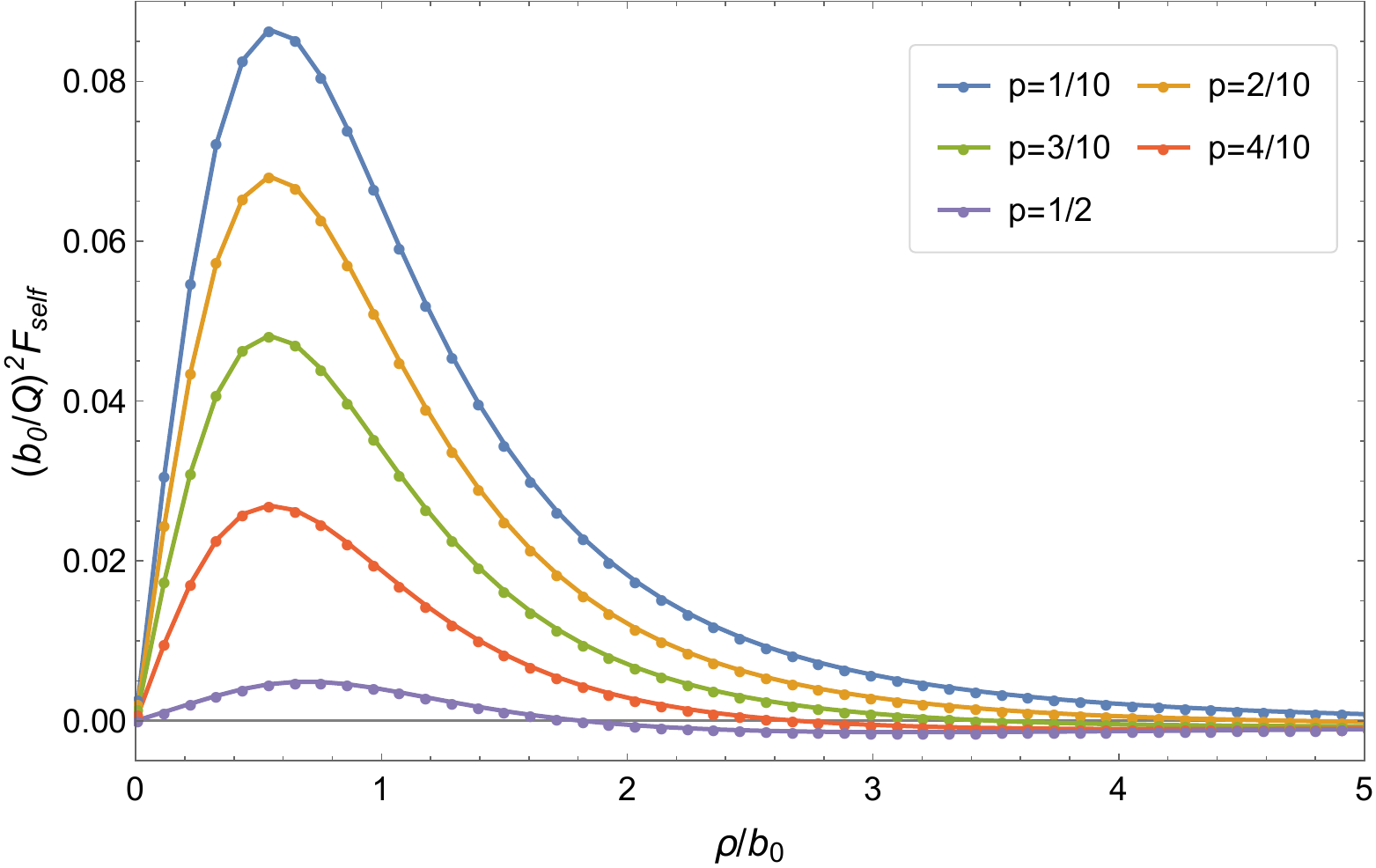}}\quad
  \subfigure[\,\,Log-plot of the absolute value of self-force for $0 < p\leq 1/2$ in (a) \label{fig:sf case1-1}]{\includegraphics[scale=0.33]{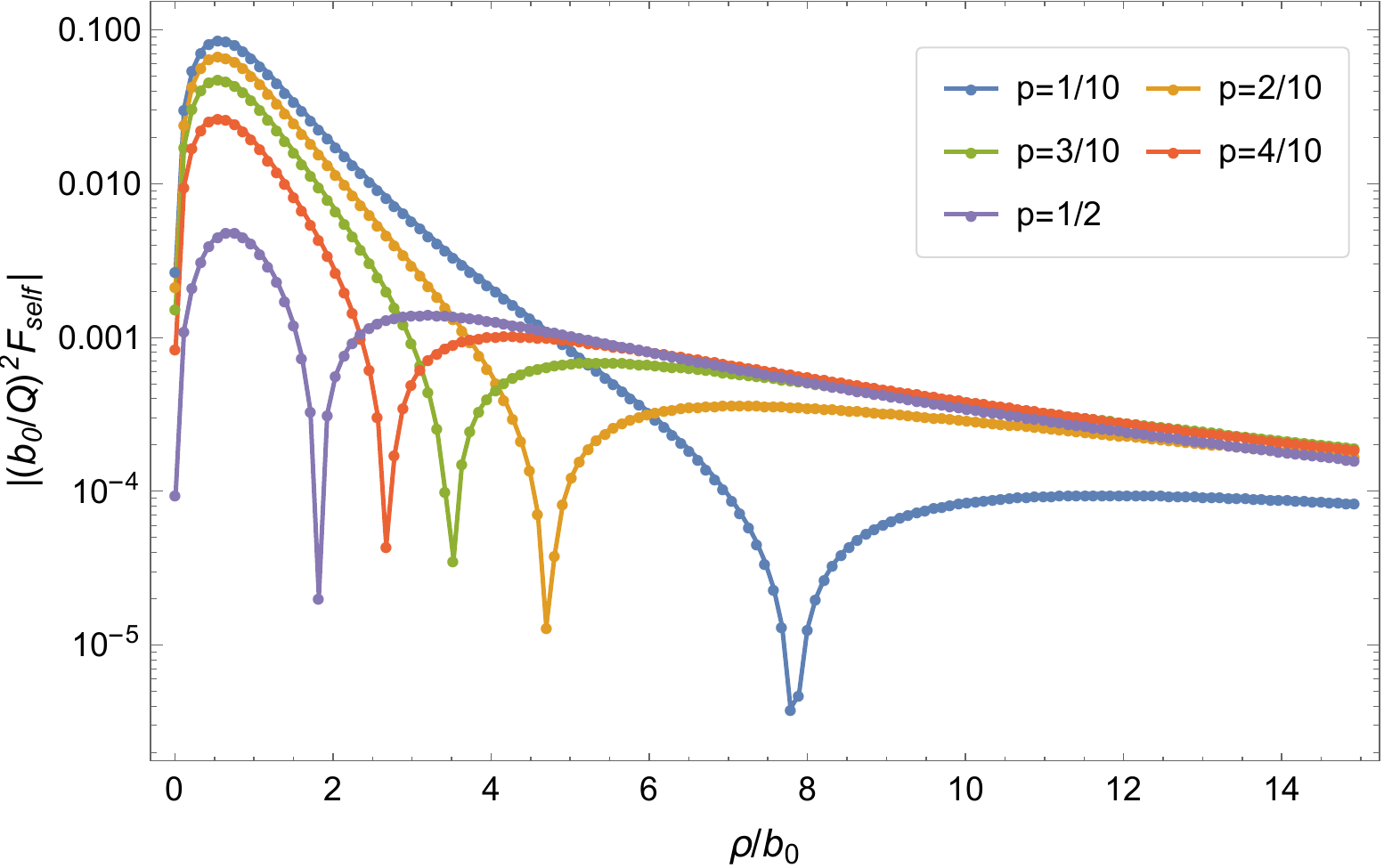}}\quad
   \subfigure[\,\,$1/2 \lesssim p \lesssim 1\,\, (q=-1,\sigma = +1)$\label{fig:sf case2}]{\includegraphics[scale=0.33]{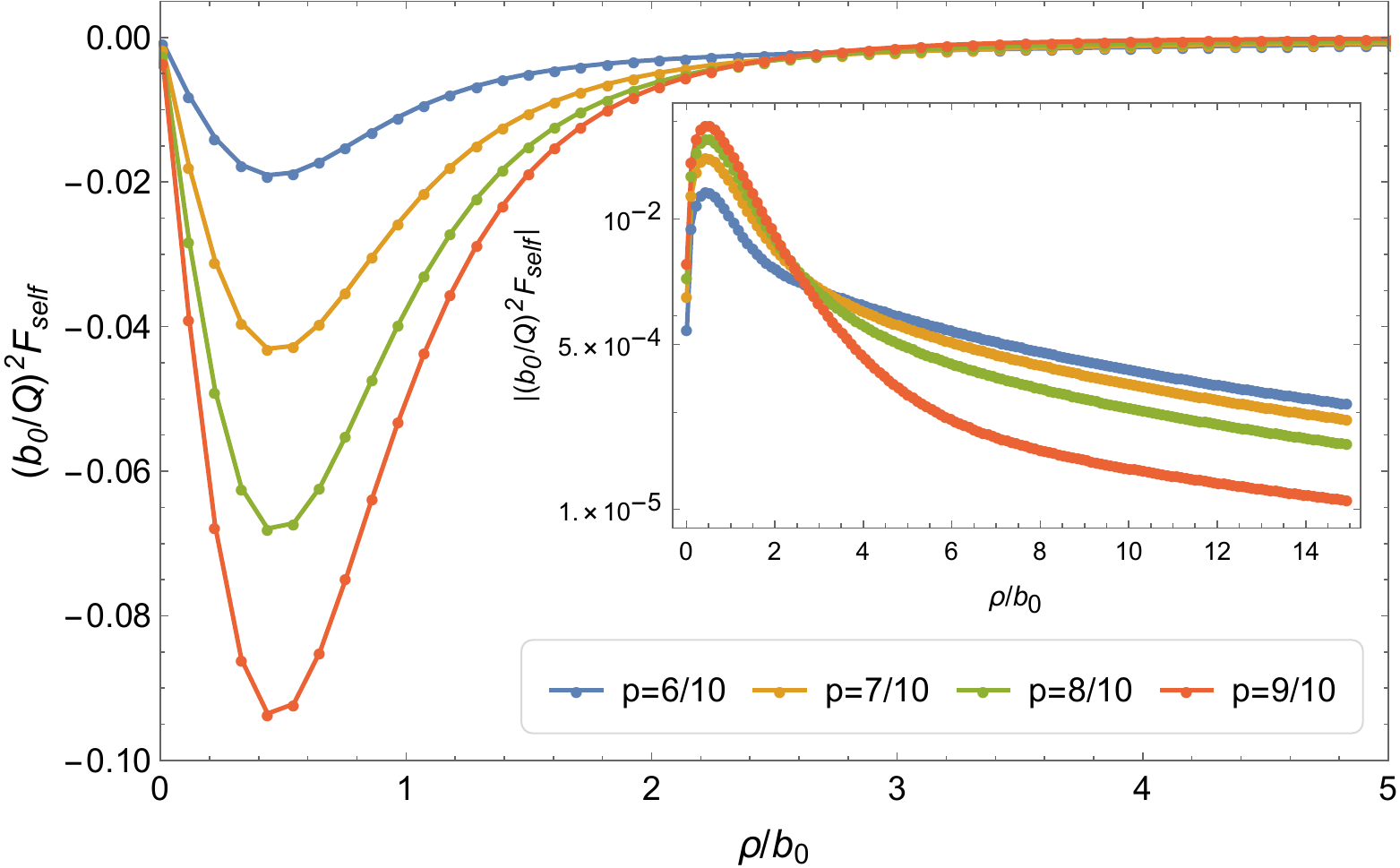}}\quad
  \subfigure[\,\,$p\geq 1 \,\, (q=-1,\sigma = +1)$\label{fig:sf case3}]{\includegraphics[scale=0.33]{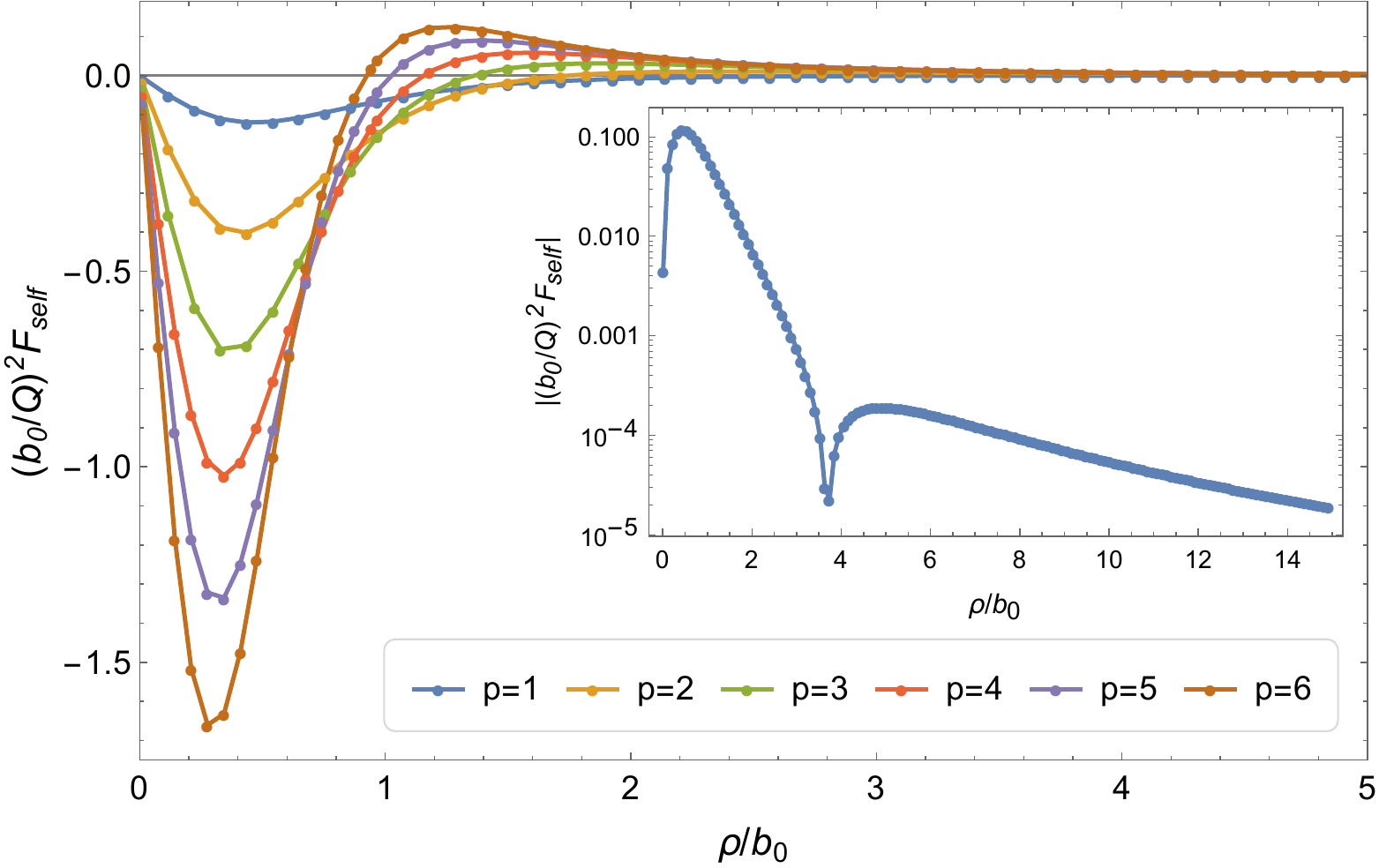}}\quad
  \subfigure[\,\,$0<p \lesssim 5 \,\, (q=-1,\sigma = -1)$\label{fig:sf case4}]{\includegraphics[scale=0.325]{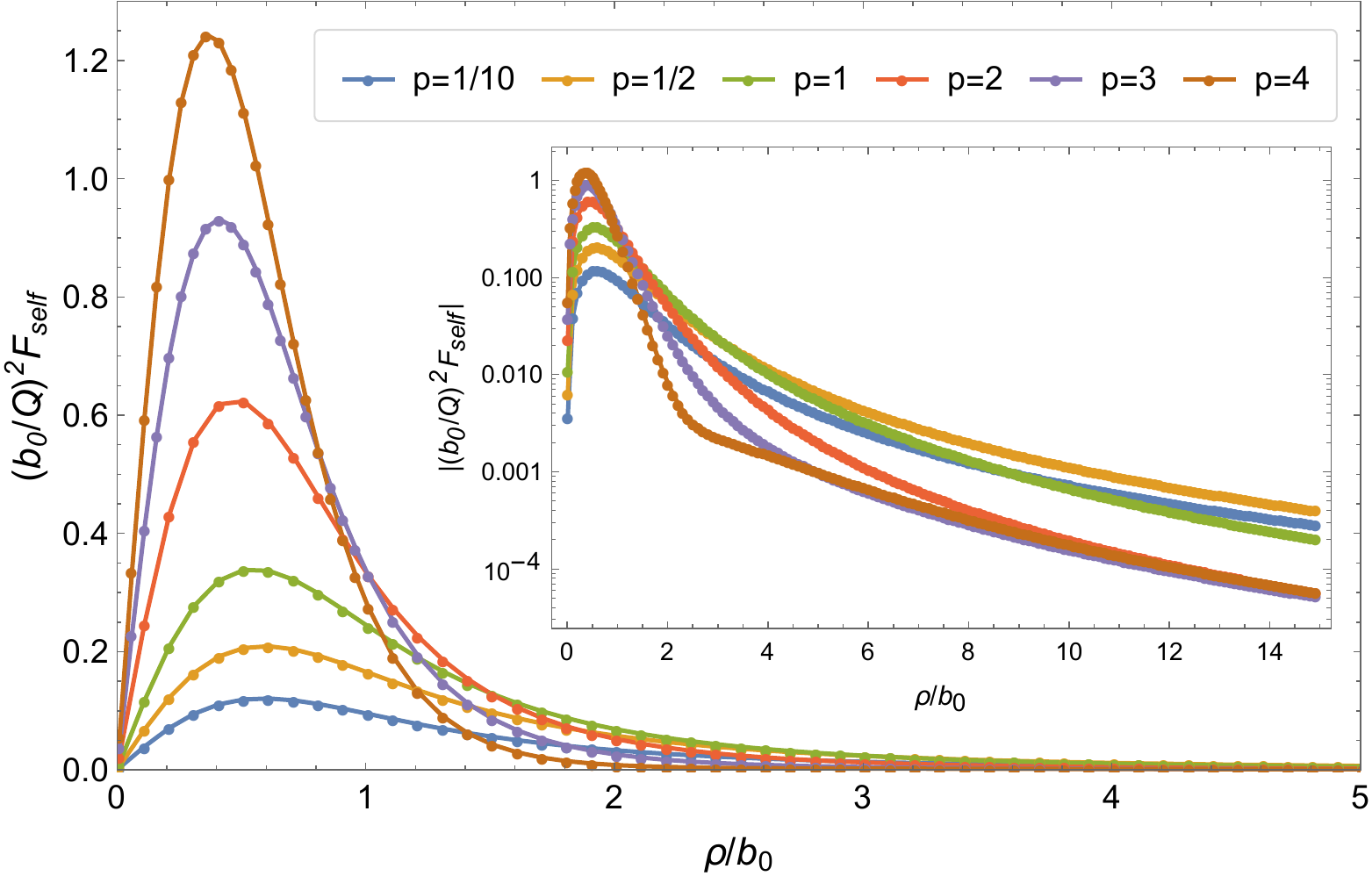}}\quad
  \subfigure[\,\,$p\geq 5\,\, (q=-1,\sigma = -1)$\label{fig:sf case5}]{\includegraphics[scale=0.325]{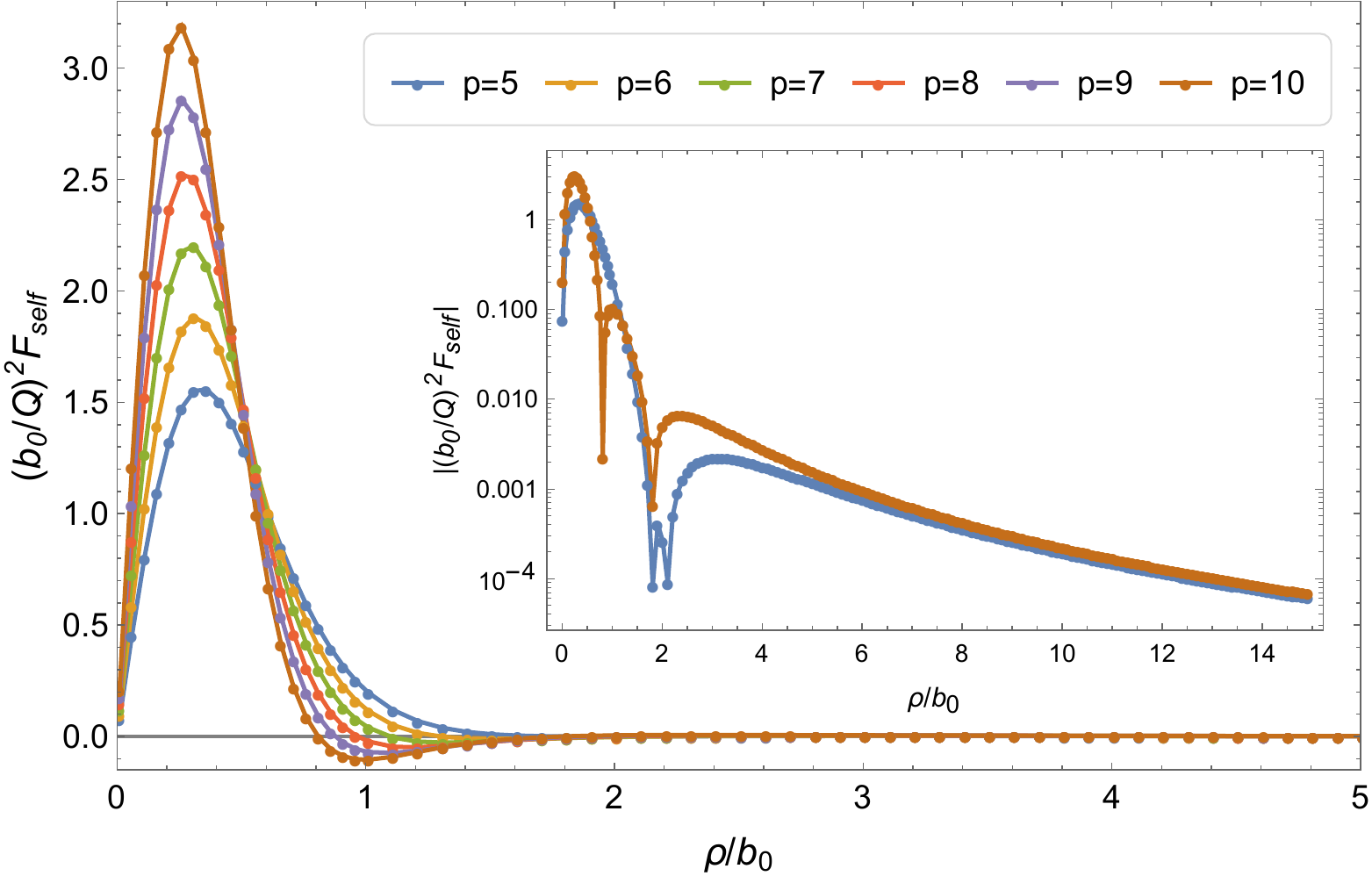}}
  \caption{Behavior of scalar self-force on a static particle in the vicinity of a wormhole with shape exponent $q=-1$ and varying parameter $p$ of the redshift function $\Phi=\sigma/r^{p}$ for (a)-(d) $\sigma=+1$, and (e)-(f) $\sigma=-1$. The insets all show the absolute value of the self-force on a log scale to make sign-switching more visible. In (d) and (f), only $p=1$ and $p=5$ are plotted in the insets because the other curves can already be clearly seen to be sign-switching.}
  \label{fig:tpneg1}
\end{figure*}

With the correct falloff obtained, we sum the regularized $l$-modes to obtain the full self-force. The general behavior of the self-force versus $\rho$ for the ultrastatic case is illustrated in Fig.~\ref{fig:sf ultrastatic} (green dots), where we emphasize that the error bars for the numerical self-force are smaller than the thickness of the points. The numerical self-force matches the exact result obtained by Taylor (red line) with a relative error given by $\varepsilon=(F_{\text {analytic }}-F_{\text {mode-sum }})/F_{\text {analytic }}$ shown as inset in Fig.~\ref{fig:sf ultrastatic}. In the presence of a static scalar charge close to the wormhole throat, the self-force grows from zero and peaks to a positive value at a certain distance near the throat, falls off while approaching zero for very large distances. This agrees with the previously obtained result that the self-force is always positive, representing a repulsion from the wormhole.

The spectrum of results is richer in the case where the gravitational redshift parameter $p$ is nonzero. 

\begin{figure*}[htp]
  \centering
  \subfigure[\,\,$0<p\lesssim 6$\label{fig:sf case11}]{\includegraphics[scale=0.25]{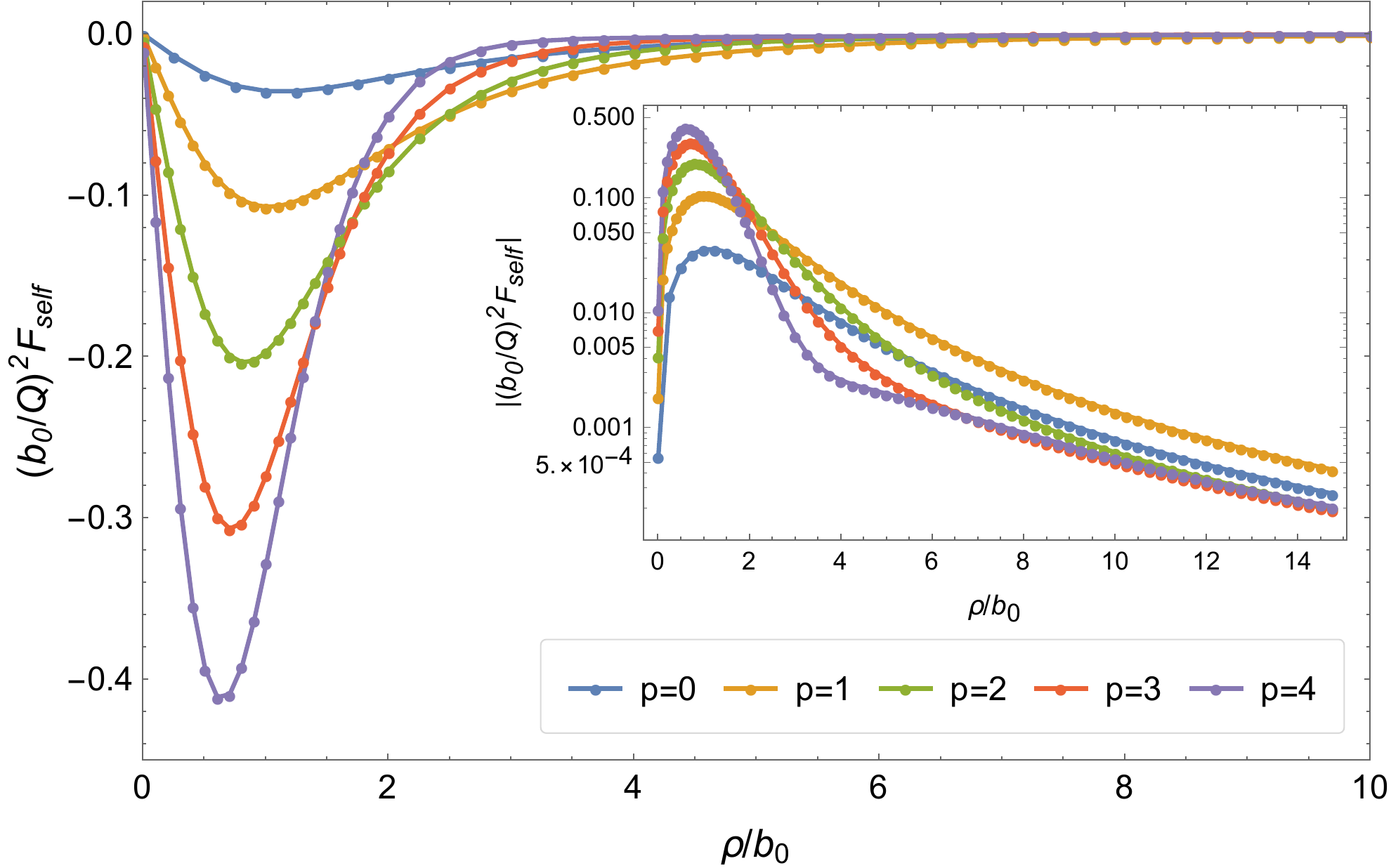}}\quad
  \subfigure[\,\,$p=5$\label{fig:sf case22}]{\includegraphics[scale=0.25]{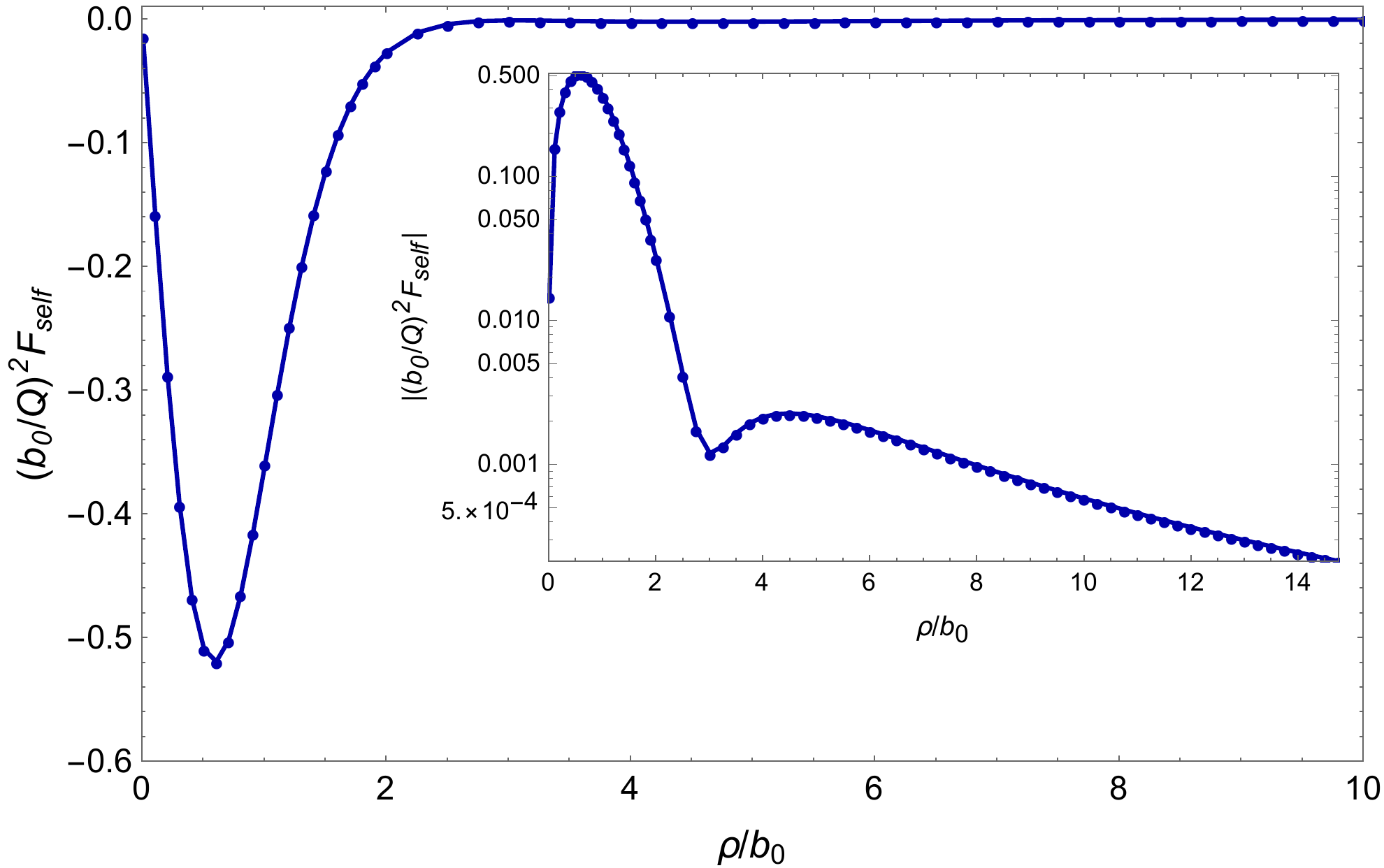}}\quad
  \subfigure[\,\,$p\geq 6$\label{fig:sf case33}]{\includegraphics[scale=0.25]{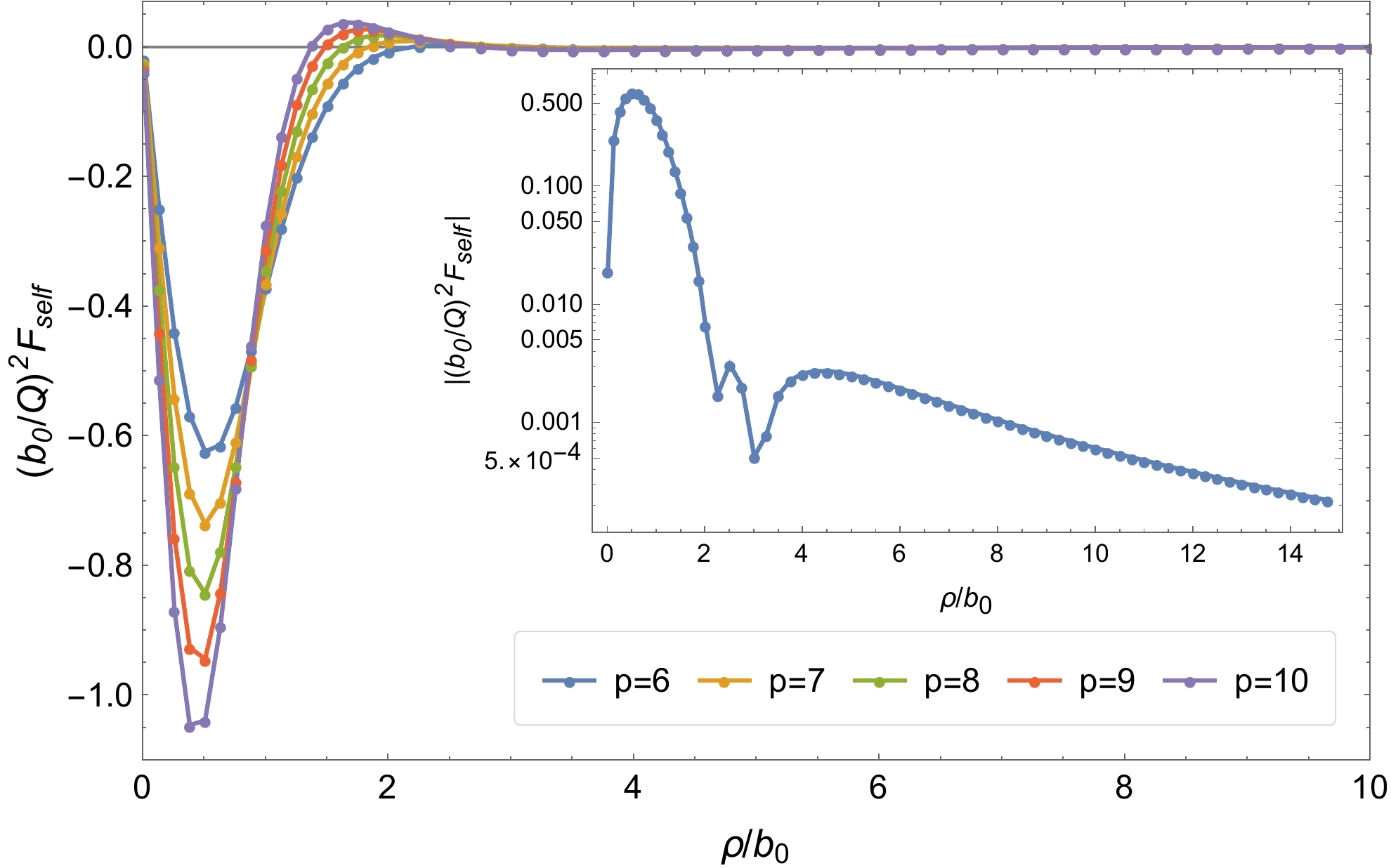}}\quad
  \subfigure[\,\,$0 < p\leq 1/2$\label{fig:sf case44}]{\includegraphics[scale=0.32]{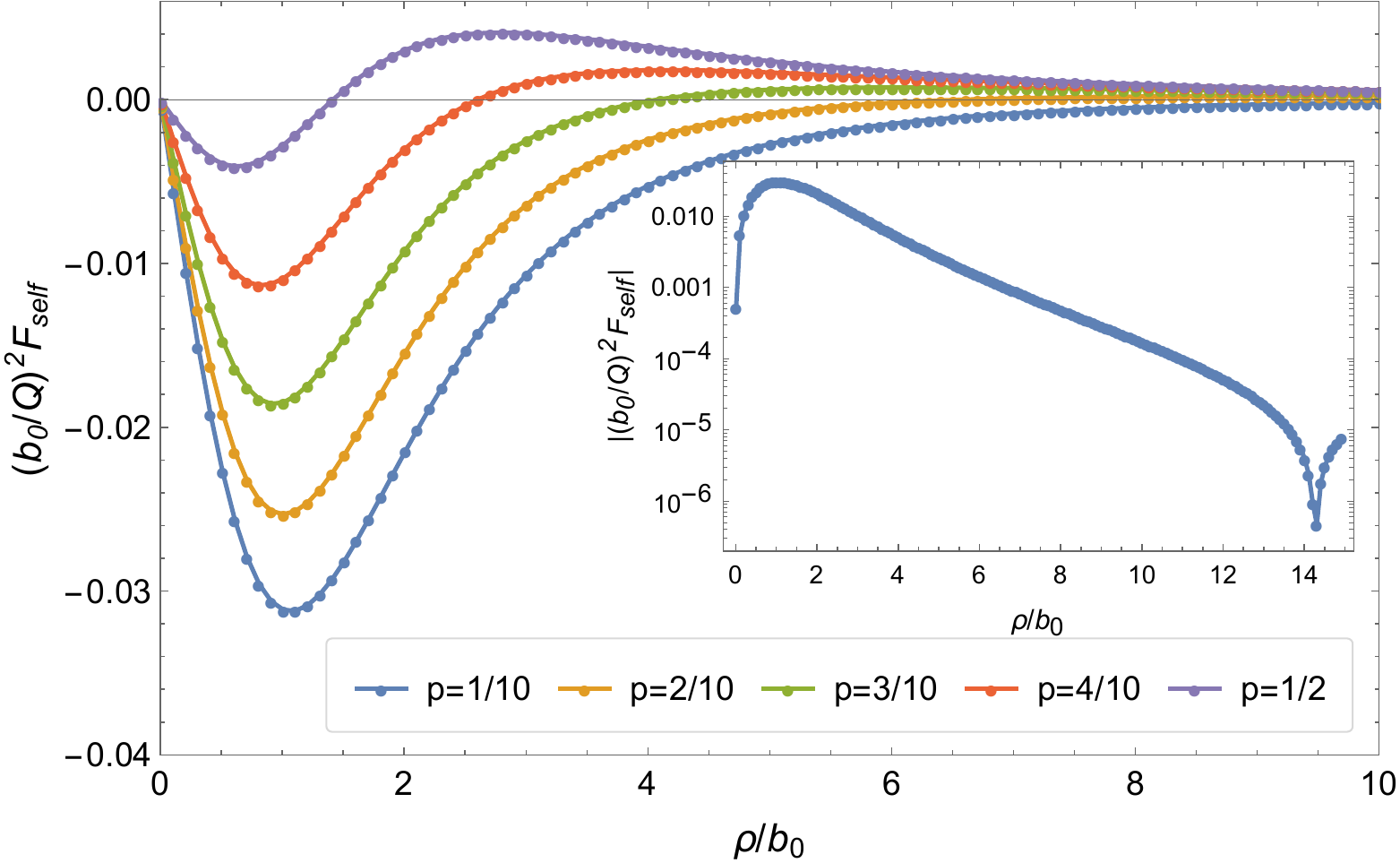}}\quad
  \subfigure[\,\,$1/2 \lesssim p \lesssim 1$\label{fig:sf case55}]{\includegraphics[scale=0.32]{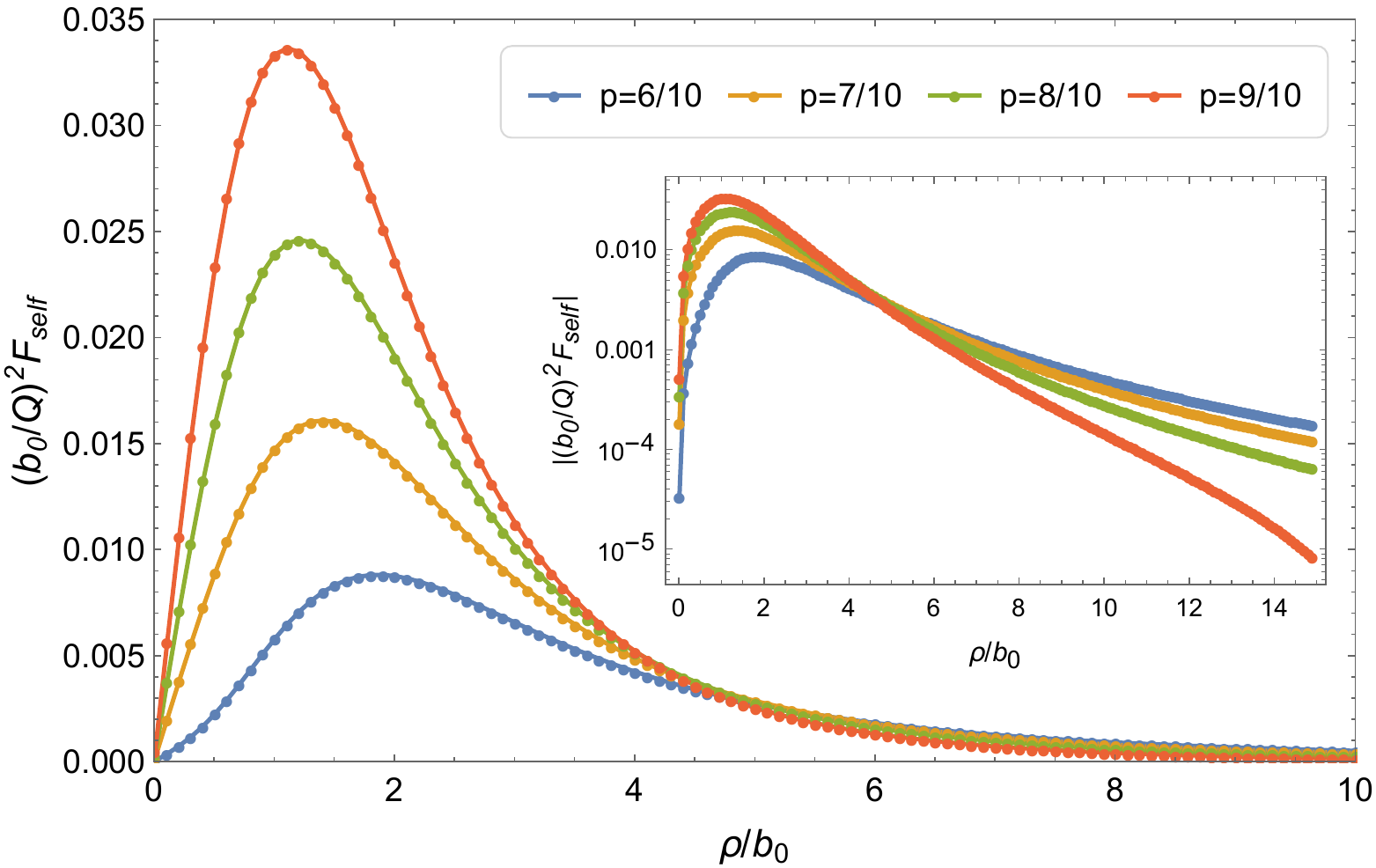}}\quad
  \subfigure[\,\,$p\geq1$\label{fig:sf case66}]{\includegraphics[scale=0.31]{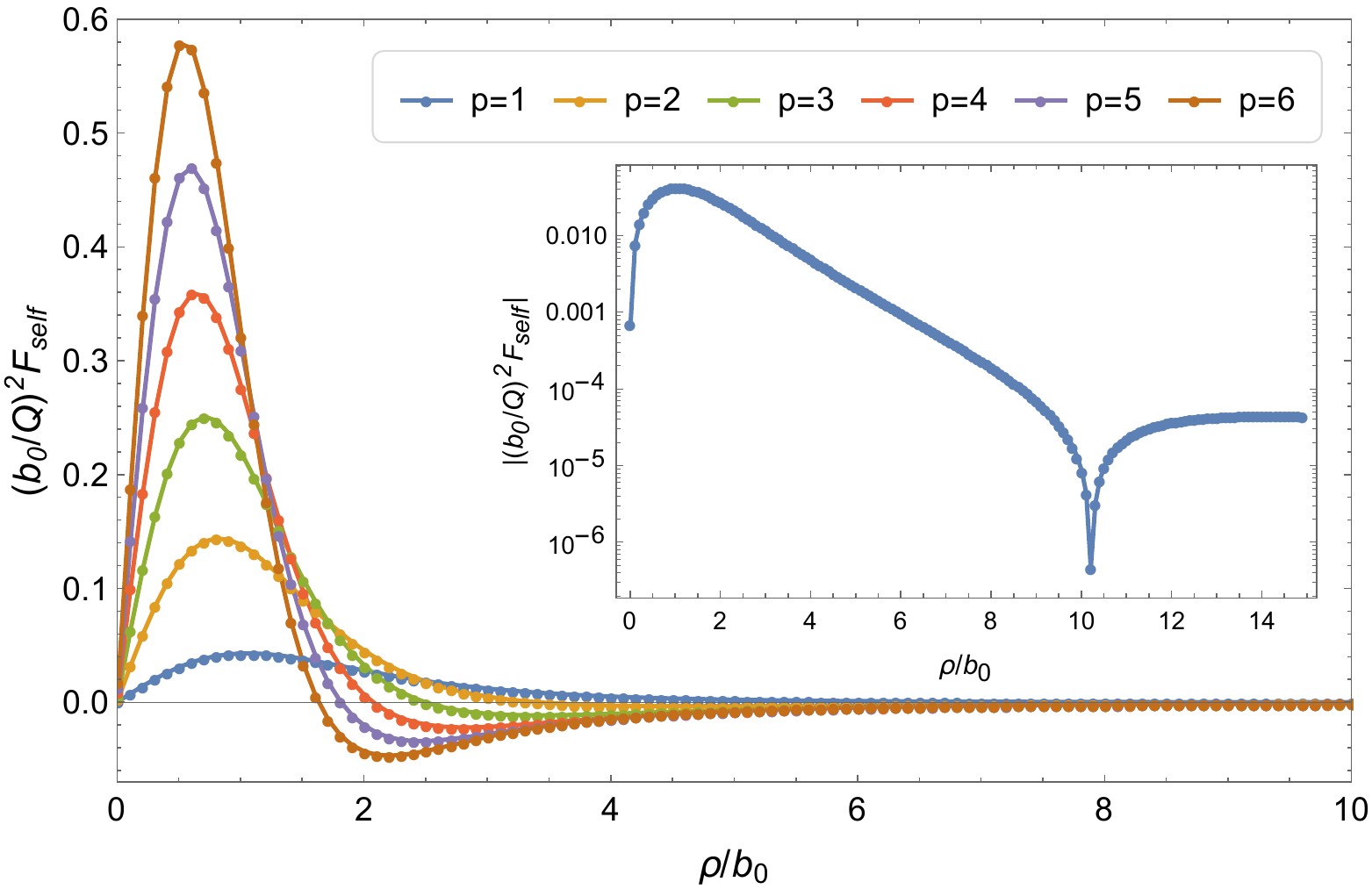}}
  \caption{Behavior of scalar self-force on a static particle in the vicinity of a wormhole with shape exponent $q=1/3$ and varying parameter $p$ of the redshift function $\Phi=\sigma/r^{p}$ for (a-c) $\sigma=+1$, and (d-f) $\sigma=-1$.}
  \label{fig:tp1over3}
\end{figure*}

\subsubsection{$q=-1, \sigma=+1$ (repulsive wormholes)}

$(a)$ $0 < p \leq 1/2$. Self-force results for various $p$ are shown in Fig.~\ref{fig:sf case1}. At first glance, the self-force appears to be everywhere positive (outward/repulsive) for all values of $p$ in this range. A closer look, however, reveals unusual behavior that is not apparent in Fig. \ref{fig:sf case1}--- the force is not always repulsive. To see this, a log-plot of the absolute value of the self-force is presented in Fig.~\ref{fig:sf case1-1}.
Apart from throat and spatial infinity, there exists a crossing point $y=y_{c}$ 
in the wormhole spacetime where the self-force vanishes. The self-force is repulsive near the throat and becomes attractive beyond the crossing point. This crossing point moves closer to the throat as $p$ increases. In the ultrastatic limit $p\rightarrow 0$, the crossing point goes to infinity and the self-force becomes repulsive everywhere, as would be expected from the Ellis wormhole. The maximum strength of the attractive self-force also appears to grow with increasing $p$. This behavior is the reverse of what happens in the repulsive region, where a larger $p$ means a weaker self-force. 

The self-force peaks at a location close to the throat, while it vanishes at the throat ($y=\rho/b_{0}=0$), and it approaches zero as $y\rightarrow\infty$. Since $\sigma=+1$, the wormhole is repulsive; to remain at a fixed position, an inward-pointing external force needs to be exerted on the particle. An outward self-force means that this required inward external force will have to be greater for a charged particle than for a neutral one. 


$(b)$ $1/2 \lesssim p \lesssim 1$. 
Slightly above $p=1/2$, the self-force is negative (i.e., inward-pointing) everywhere (see Fig.~\ref{fig:sf case2}). The inset that displays the logarithm of the absolute value of the self-force makes this observation clear. Much like the attractive branch of the previous case (i.e., $0 < p\leq 1/2$), the strength of the self-force increases with increasing $p$.


$(c)$ $p\geq 1$. The behavior of the self-force for the range $p\geq 1$ is evident even in Fig.~\ref{fig:sf case3}. We observe the existence of self-force crossings for all $p$ values used. Close to the wormhole, the self-force points inward, but it points outward beyond a certain distance. Same as the $1/2\lesssim p\lesssim 1$ case, an increase (decrease) in $p$ increases (decreases) the maximum strength of the self-force.



\begin{table*}[tbp]
\centering
\begin{threeparttable}
\begin{tabular}{c@{\hskip 1cm}c@{\hskip 1cm}c}
\hline \hline
	\textbf{shape exponent} & \textbf{blueshift ($\sigma=+1$)} & \textbf{redshift ($\sigma=-1$)}\\ 
	\hline 
	$q=-1$ & \makecell{$0<p\leq 1/2$: with crossing, decreasing \textsuperscript{i} \\ $1/2 \lesssim p \lesssim 1$: attractive, increasing \\ $p\geq 1$: with crossing, increasing}  & \makecell{ $0<p\lesssim 5$: repulsive, increasing \\ $p\geq 5$: with crossing, increasing}\\ [0.4cm]
$q=1/3$ & \makecell{ $0<p\lesssim 6$: attractive, increasing \\ $p\geq 6$: with crossing, increasing}  & \makecell{$0<p\leq 1/2$: with crossing, decreasing \\ $1/2\lesssim p\lesssim 1$: repulsive, increasing \\ $p\geq 1$: with crossing, increasing} \\[0.2cm] \hline \hline
\end{tabular}
\begin{tablenotes}
\item[i] The sign of the self-force is denoted by the first entry, followed by the behaviour of its magnitude in response to an increase in redshift parameter $p$.
\end{tablenotes}
\end{threeparttable}
\caption{Summary of the behaviour of self-force for the wormholes with exact throat profile and with two cases of frequency shifts; there exists a crossing in the self-force for some $p$ which signifies a change in its sign depending on the distance of the scalar charge from the wormhole throat.}
\label{tab:tablesummary}
\end{table*}

\begin{figure*}[htp]
  \centering
  \subfigure[\,\,$q=-1$ \label{fig:crossings1}]{\includegraphics[scale=0.28]{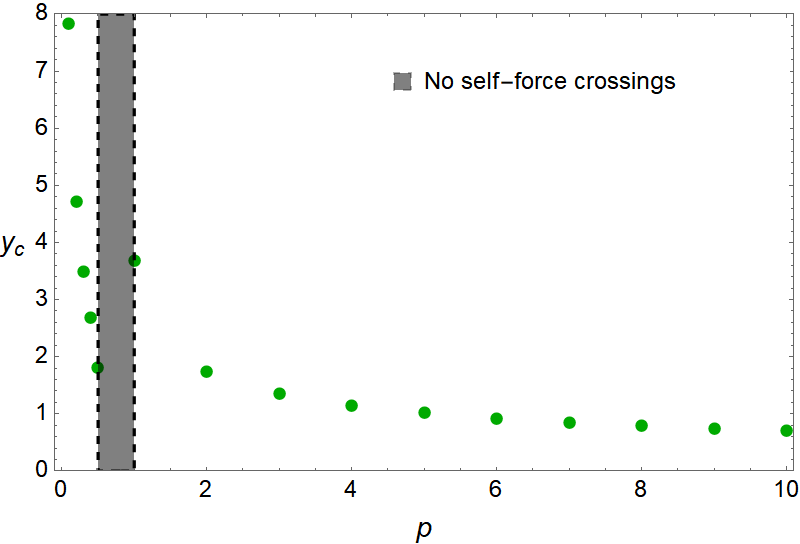}}\quad
  \subfigure[\,\,$q=1/3$ \label{fig:crossings2}]{\includegraphics[scale=0.3]{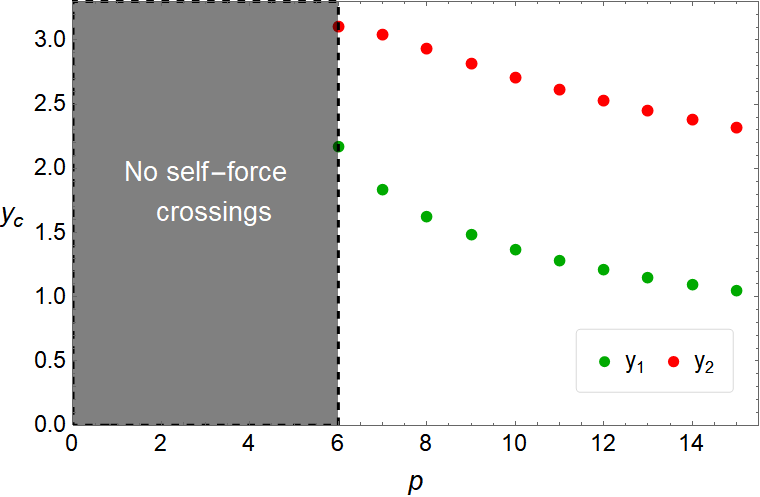}}
  \caption{Behavior of the roots of vanishing self-force as a function of $p$ for $\sigma=+1$ in the case (a)\,$q=-1$, (b)\,$q=1/3$; the same behavior holds true for $\sigma=-1$.}
  \label{fig:sf crossings}
\end{figure*}

\subsubsection{$q=-1, \sigma=-1$ (attractive wormholes)}
$(a)$ $0<p\lesssim 5$. By reversing the sign of the redshift function, the behaviour of the self-force also changes (see Figs.~\ref{fig:sf case4} and \ref{fig:sf case5}). From a previous attractive nature, the self-force in this case becomes repulsive everywhere. As $p$ increases within this range, the magnitude of the self-force also increases. Since this is an attractive wormhole, a particle requires an outward external force in order to remain at a fixed position. An outward self-force implies that this required external force will be less for a charged particle than for a neutral one.

$(b)$ $p\geq 5$. For $p=5$ and beyond, the self-force is no longer entirely repulsive. Specifically, there exist two crossing points for all $p$ in this range, i.e., the self-force starts as repulsive near the throat, then goes attractive, and finally becomes repulsive again farther away from the throat. There ought to exist a transition $p_*$ in the range $4 < p < 5$ at which sign-switching for the self-force starts, but we are unable to determine this transition.

\subsection{$(q=1/3)$ wormholes}
Wormholes belonging to this class are described by the throat profile given in Eq.~\eqref{tp1}, and seem to have been largely ignored in the literature. In terms of their embedding diagrams, these wormholes flare out slightly less drastically compared to $(q=-1)$ wormholes. (See. Fig.~\ref{fig:embedding}.) 


\subsubsection{$q=1/3, \sigma=+1$ (repulsive wormholes)}
$(a)$ $0\leq p\lesssim 6$. As shown in Fig.~\ref{fig:sf case11}, the self-force appears to be everywhere negative for all values parametrizing the gravitational redshift, and this is verified by looking at the absolute value of the self-force. This means that for a repulsive wormhole, the required external inward-directed force is less since the computed self-force is negative. For fixed $y=\rho/b_0$, the magnitude of the self-force increases with increasing $p$. It can also be observed that when the particle is situated a little bit farther from the throat (e.g., $y=2$), the behavior of self-force is different for an arbitrary value of $p$. Particularly, the self-force increases in magnitude as $p$ increases from $0$ to $2$ and is suppressed when $p$ is continuously increased from $p=3$. 

At this point, we strongly expect a transition from an attractive self-force into a repulsive one. In fact, the transition is already evident for the case $p=5$. This can be seen through the kink at $y\approx 3$ in the inset plot shown in Fig.~\ref{fig:sf case22}.

\begin{figure*}[htp]
  \centering
  \subfigure[\,\,Absolute deviation between exact and numerical results\label{fig:error check}]{\includegraphics[scale=0.32]{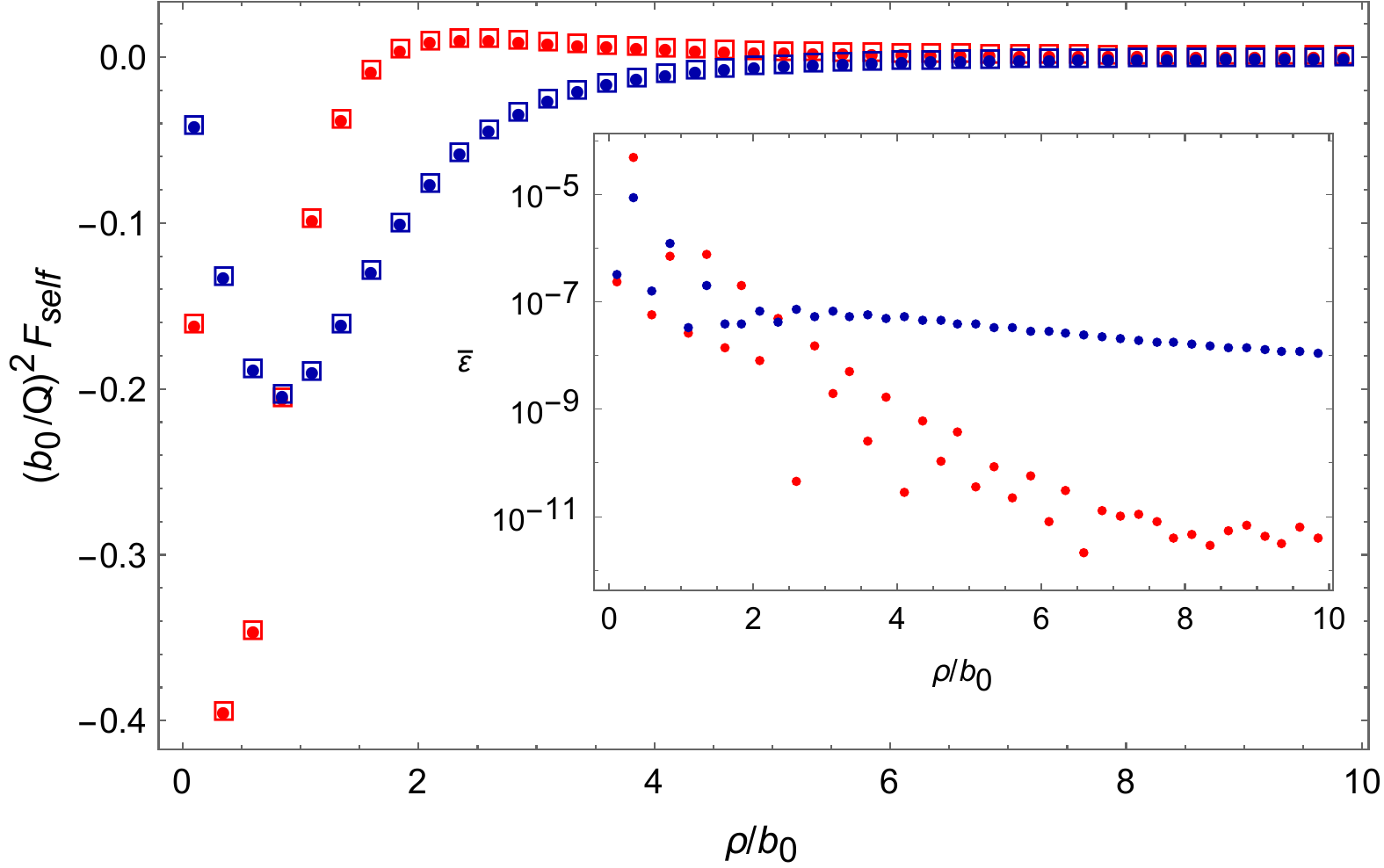}}\quad
  \subfigure[\,\,Ultrastatic wormholes ($p=0$)\label{fig:NS case1-1}]{\includegraphics[scale=0.29]{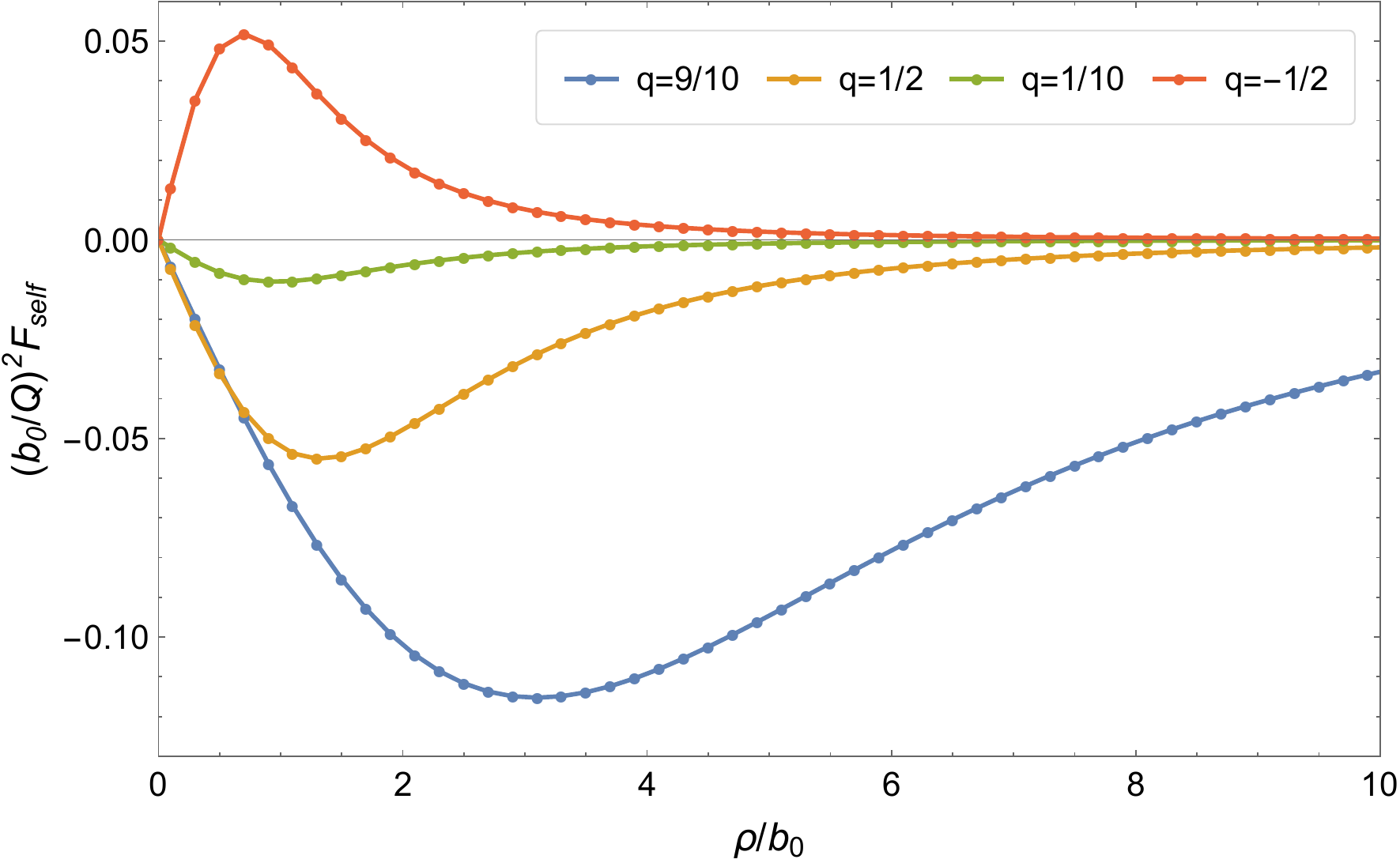}}\quad
   \subfigure[\,\,$p=1$\label{fig:NS case2}]{\includegraphics[scale=0.286]{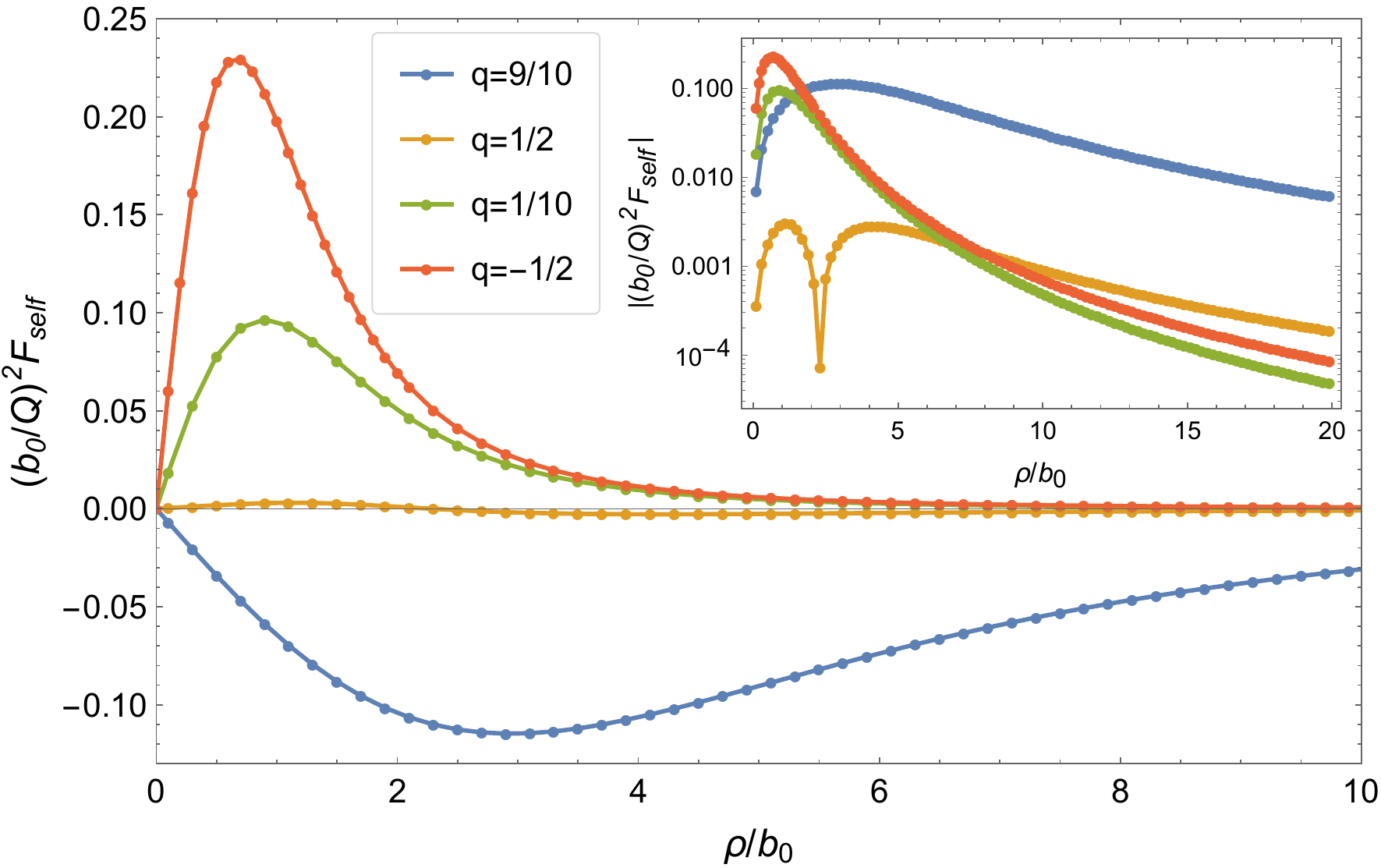}}\quad
  \subfigure[\,\,$p=5$\label{fig:NS case3}]{\includegraphics[scale=0.282]{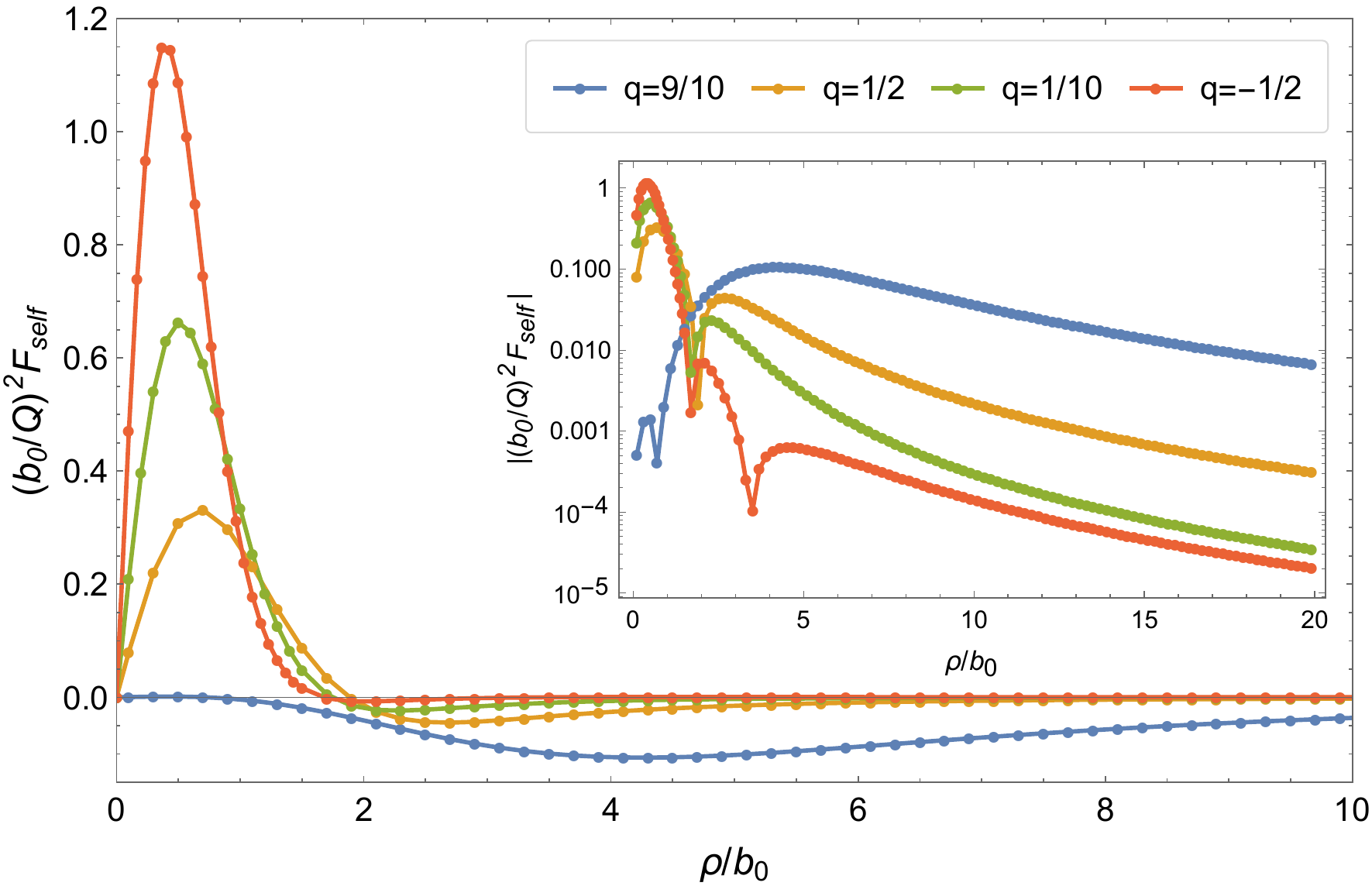}}
  \caption{In (a), the red and blue markers represent the self-force obtained for $(q=-1, p=2)$ and $(q=1/3, p=2)$, respectively. The value of the self-force computed using the exact throat profile is denoted by square markers, while the filled circles are for values obtained from the numerical code for generic $(q, p)$; the inset in (a) shows the log-plot of the absolute deviation between these two values: $\bar{\varepsilon}=|F_{exact}-F_{numerical}|$. In (b)-(d), the plots reveal the behavior of scalar self-force on a static particle in the vicinity of a numerical wormhole with a fixed parameter $p$ of the redshift function $\Phi=-1/r^{p}$, and varying shape exponent $q$.}
  \label{fig:numerical}
\end{figure*}

$(b)$ $p\geq 6$. For $p=6$ and beyond, we produce distinct features of scalar self-force with respect to the first throat profile considered in this work. As presented in Fig.~\ref{fig:sf case33}, we also observe the basic qualitative feature seen for the case $q=-1$, which is essentially the presence of self-force crossings. Near the throat, the self-force is attractive for all values of $p$ with a magnitude that gets amplified with increasing $p$. Farther away from the throat, the self-force becomes repulsive, and it persists even in an approximate infinity. Initially, this is the observation from generating a simple self-force plot, but if we look closer, then a more interesting feature is observed. An inset plot in Fig.~\ref{fig:sf case33} explicitly shows that the self-force curve crosses the zero point twice. The self-force acting on the scalar particle undergoes a transition from being attractive to repulsive and back to attractive. The two crossings tend to move closer to the wormhole throat with a stronger blueshift effect (i.e., increasing $p$), as shown in Fig.~\ref{fig:sf crossings}. This particular self-force behavior resembles a sort of ``damped oscillation''--- the oscillatory behavior fades away with distance from the wormhole throat. Depending on the position of the particle (hence the sign of the self-force), for these particular repulsive wormholes, the direction of the required external force needs to be adjusted accordingly in order to keep the particle at rest.

\subsubsection{$q=1/3, \sigma=-1$ (attractive wormholes)}
$(a)$ $0<p\leq 1/2$.  We present here the behavior of the self-force when the redshift function is described by $\Phi=-1/r^{p}$, thereby representing attractive wormholes(see Figs.~\ref{fig:sf case44}-\ref{fig:sf case66}). The self-force shifts from attractive to repulsive as the particle is situated farther from the throat. As $p$ increases, the magnitude of the attractive self-force decreases while its repulsive (tail) part increases.

$(b)$ $1/2\lesssim p\lesssim 1$. For this range of $p$ values, the computed self-force becomes purely repulsive with increasing magnitude as the redshift parameter increases. This means that for an attractive wormhole, the outward-directed external force required to hold the particle in place is less for a charged particle than for a neutral one.

$(c)$ $p\geq1$. From purely repulsive self-force, a further increase in $p$ generates an attractive (tail) part. The magnitude of the self-force is amplified by increasing $p$ while the crossing points generally move towards the wormhole throat (see Fig.~\ref{fig:sf crossings}). 

For a summary of the qualitative features of the computed self-force for the two wormholes with exact throat profile, please see Table~\ref{tab:tablesummary}.

\begin{figure*}[htp]
  \centering
  \subfigure[\,\,$\sigma=+1$\label{fig:surfacep}]{\includegraphics[scale=0.36]{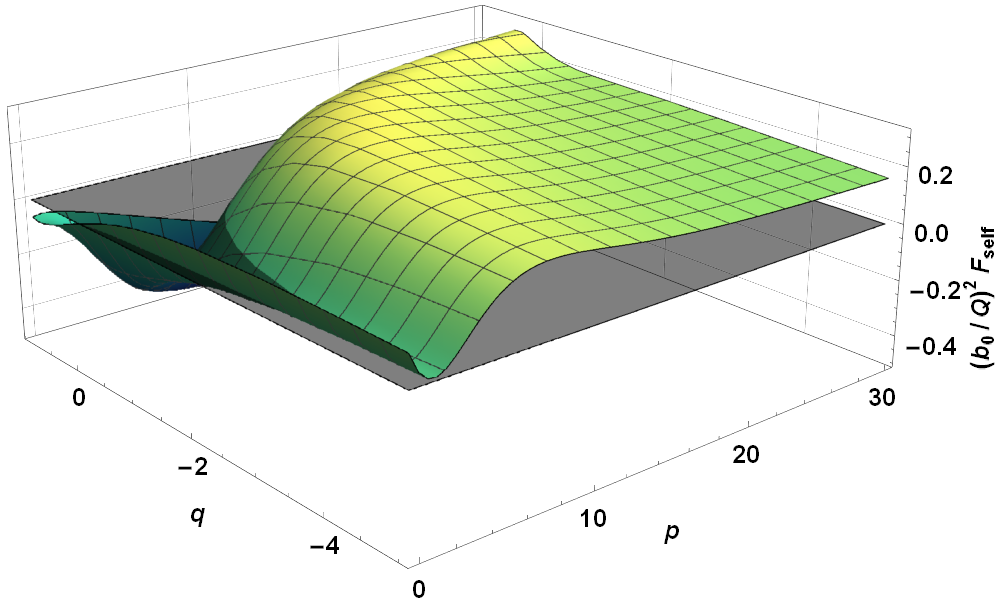}}\quad
  \subfigure[\,\,$\sigma=+1$\label{fig:densityp}]{\includegraphics[scale=0.28]{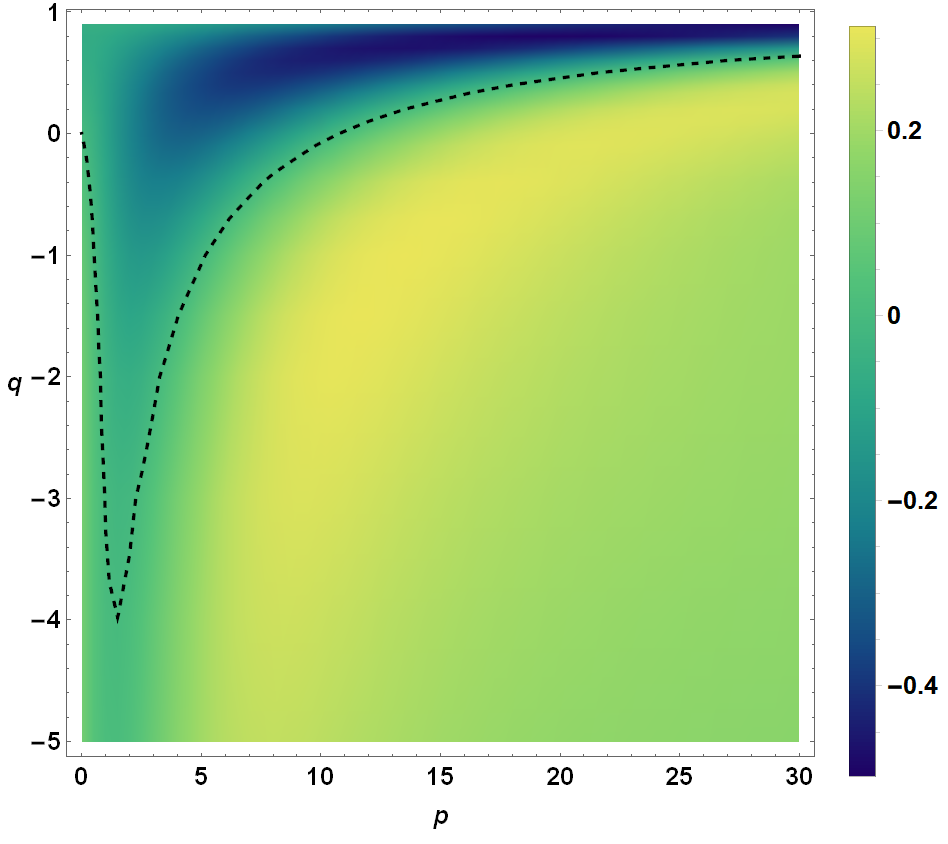}}\quad
  \subfigure[\,\,$\sigma=-1$\label{fig:surfacen}]{\includegraphics[scale=0.36]{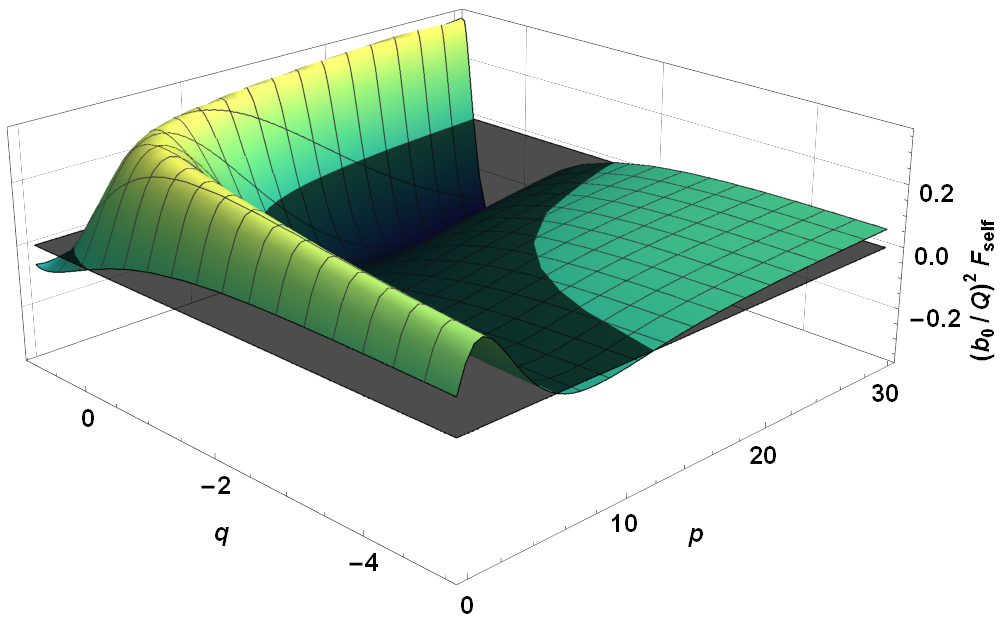}}\quad
  \subfigure[\,\,$\sigma=-1$\label{fig:densityn}]{\includegraphics[scale=0.28]{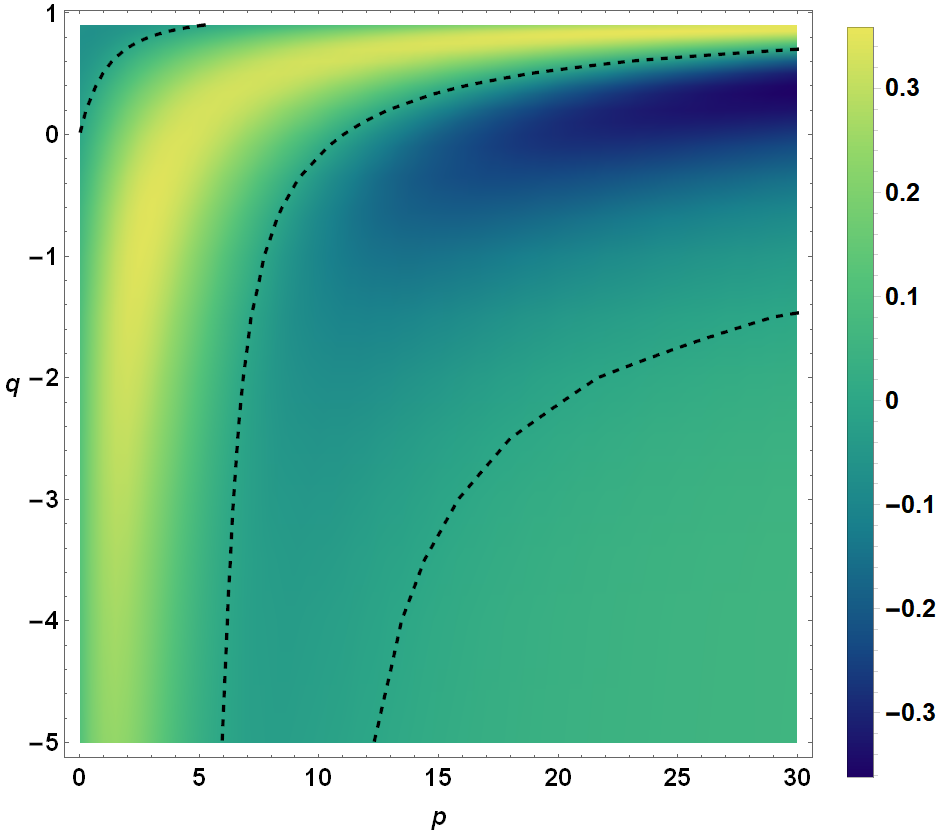}}
  \caption{Three-dimensional plot with corresponding density plot of the self-force on a static scalar charge in the Konoplya-Zhidenko two-parameter family of traversable wormholes, particle at $y=\rho/b_{0}=1$; surface plot and dashed lines in black correspond to vanishing self-force.}
  \label{fig:mainresult}
\end{figure*}

\subsection{Numerical wormhole spacetimes}
\label{subsec:num}

We present in Fig.~\ref{fig:numerical} the self-force on a static scalar charge in a set of wormhole spacetimes defined by numerical throat profiles parametrized by $q$, and with constant redshift parameter $p$. In Fig.~\ref{fig:error check}, the absolute error shows the absolute value of the difference between the self-force obtained using the analytic throat profile and via the numerical code for a generic KZ wormhole. For the cases considered, $(q=-1, p=2)$ and $(q=1/3, p=2)$, the maximum error amounts to $10^{-5}$. This demonstration gives us enough confidence that our implementation of the mode-sum regularization procedure proves to be successful in self-force calculations involving spacetimes defined by numerical metric functions, which we claim as a first in the literature. 

\subsubsection{Ultrastatic wormholes}
In the previously considered cases of ultrastatic wormholes with shape exponent $q=-1$ and $q=1/3$, the self-force computed is found to be purely repulsive and attractive, respectively. Given this, we speculate that no ultrastatic wormhole produces a self-force profile with crossings. By examining other throat profiles, we verify that this is actually the case, as shown in Fig.~\ref{fig:NS case1-1}. The self-force changes direction from an absolute attractive to repulsive as we consider more negative values of $q$, in which the wormhole geometry flares out more.

\subsubsection{Generic $(q,p)$ wormholes}
As $p$ is increased to unity, self-force crossing starts to appear again in Fig.~\ref{fig:NS case2} for $q=1/2$. The self-force acting on the scalar charge in the vicinity of the throat is repulsive and becomes attractive as the charge is placed farther away, while its magnitude increases with a more negative $q$ value. In Fig.~\ref{fig:NS case3}, for the case $q=-1/2$, the wormhole gives rise to at least two crossing points or two transitions in the computed self-force.

Given the large class of cases considered in this work, our implementation of the mode-sum approach is now expected to be capable of performing self-force calculations involving a static scalar charge in any static spherically symmetric traversable wormhole that can be modelled by a shape exponent $q<1$ and redshift parameter $p\geq 0$. Suppose, for instance, the scalar charge is situated at a distance $y=1$ from the wormhole throat, then the resulting self-force has a profile presented in Fig. \ref{fig:mainresult} for any choice of wormhole shape and strength of redshift effect. Any point ($q$, $p$) along the black dashed lines corresponds to a vanishing self-force.

\begin{figure*}[htp]
  \centering
  \subfigure[\label{fig:beta1}]{\includegraphics[width=0.32\linewidth]{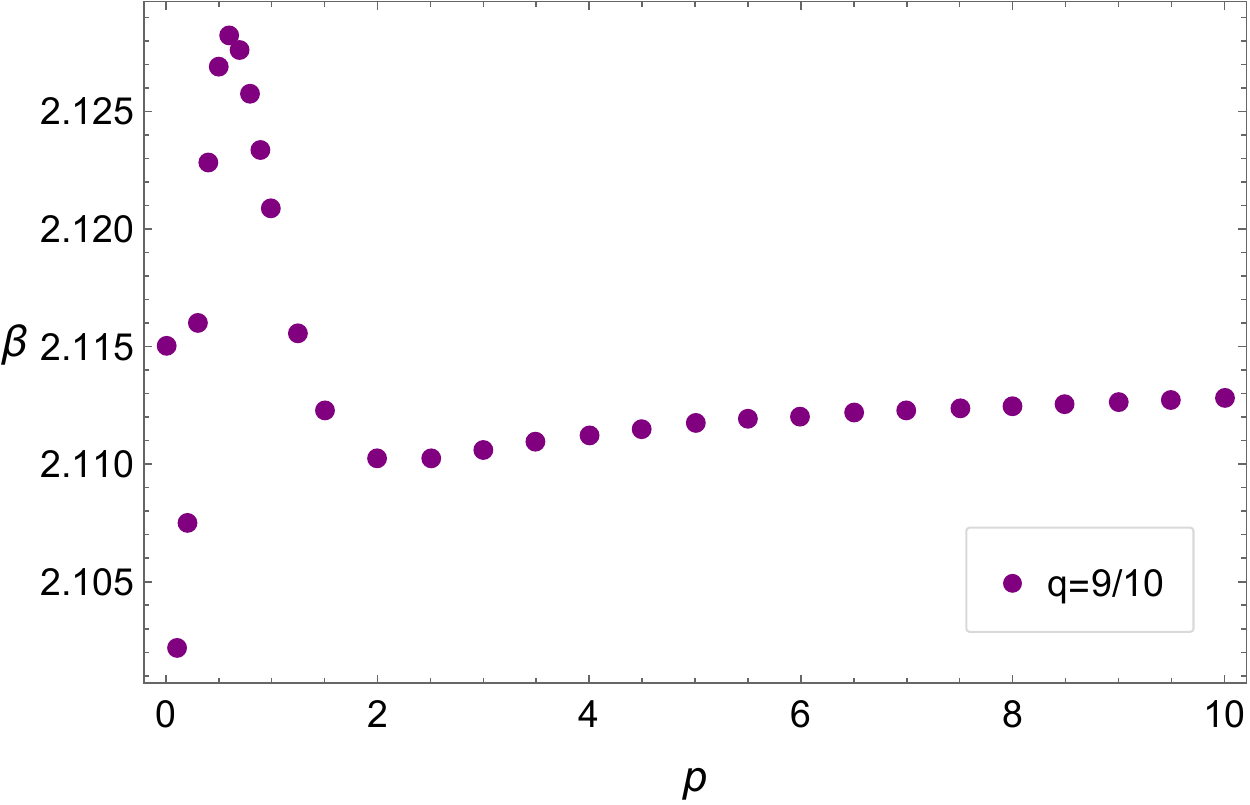}}\quad
  \subfigure[\label{fig:beta2}]{\includegraphics[width=0.32\linewidth]{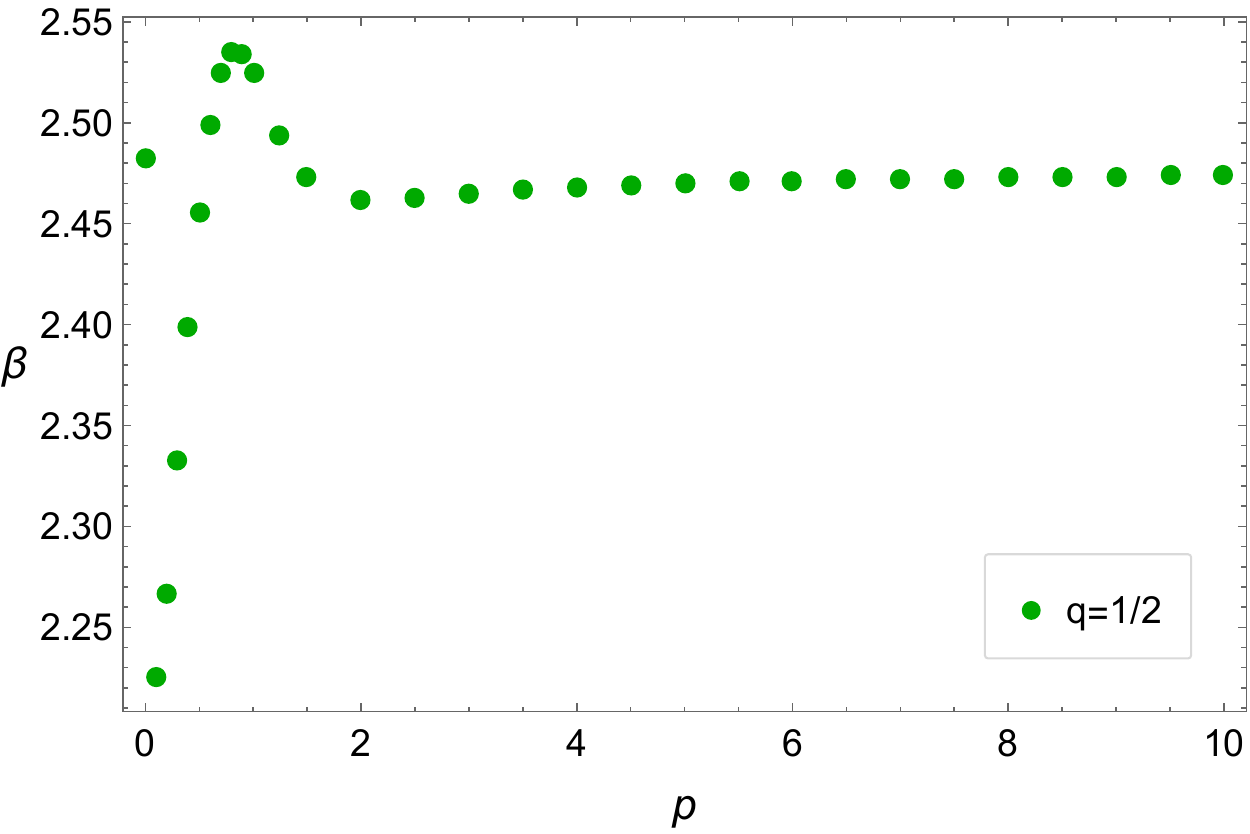}}\quad
  \subfigure[\label{fig:beta3}]{\includegraphics[width=0.32\linewidth]{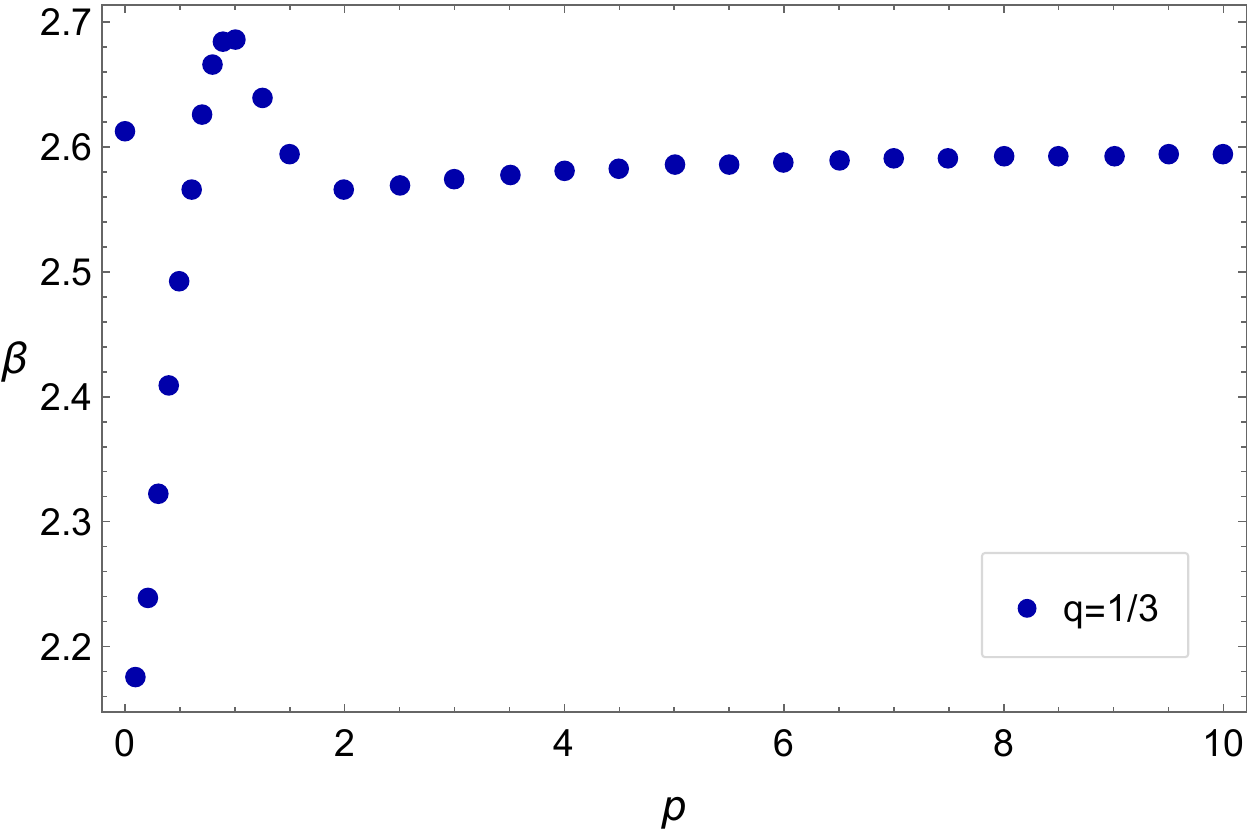}}\quad
  \subfigure[\label{fig:beta4}]{\includegraphics[width=0.32\linewidth]{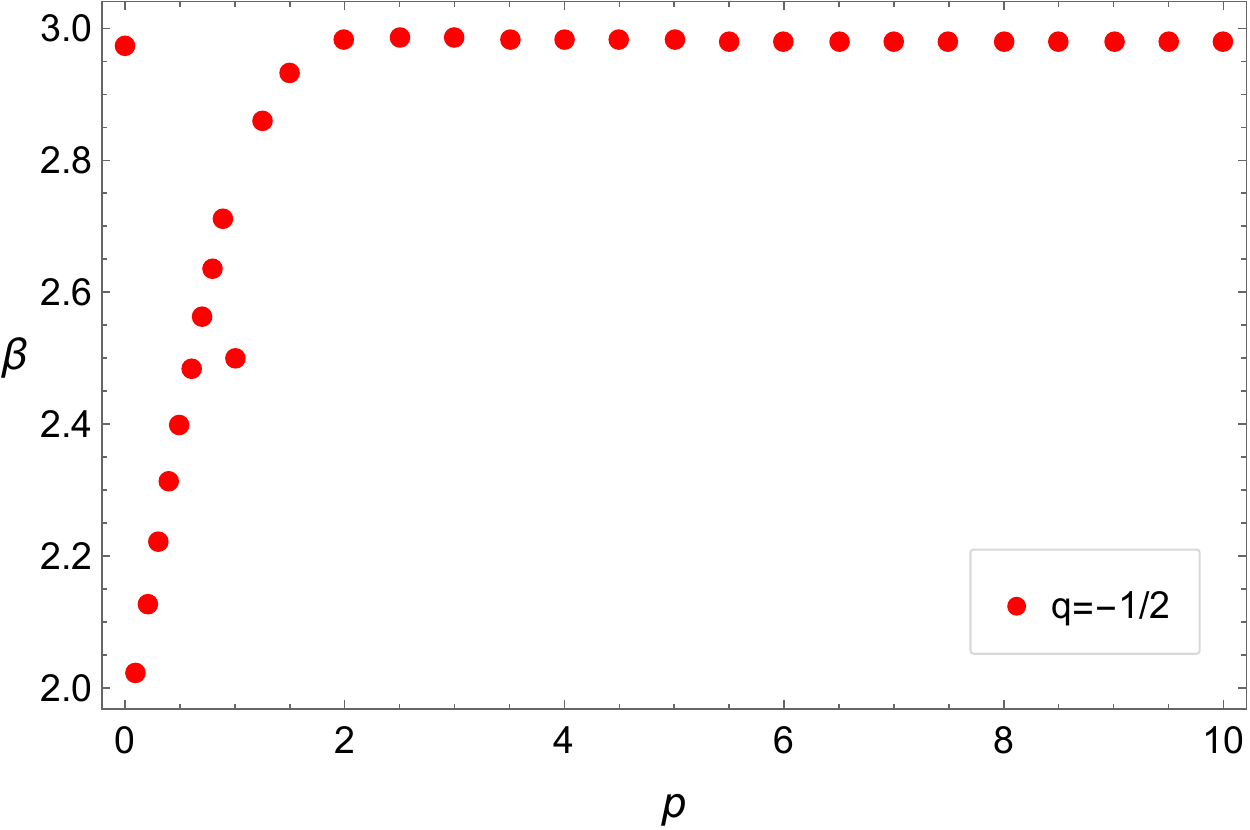}}\quad
  \subfigure[\label{fig:beta5}]{\includegraphics[width=0.32\linewidth]{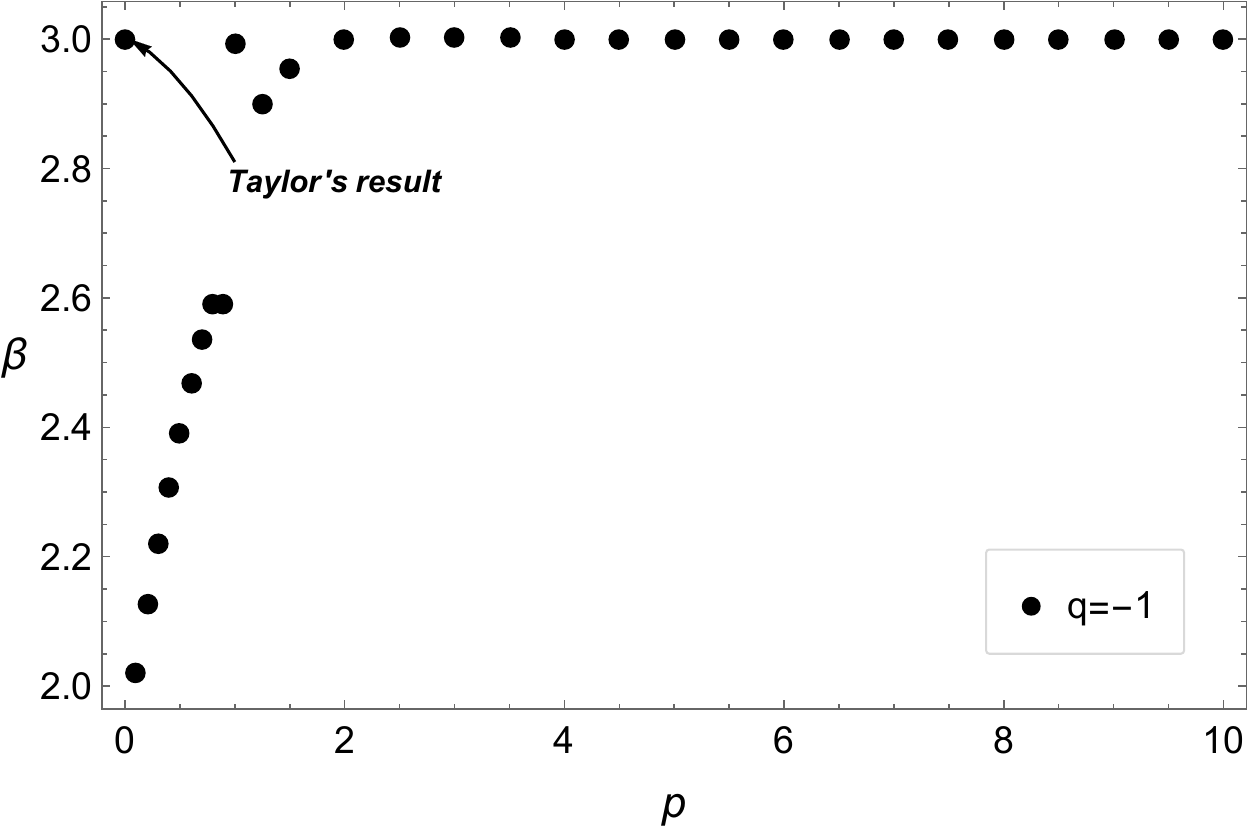}}\quad
  \subfigure[\label{fig:sf asymp q}]{\includegraphics[width=0.32\linewidth]{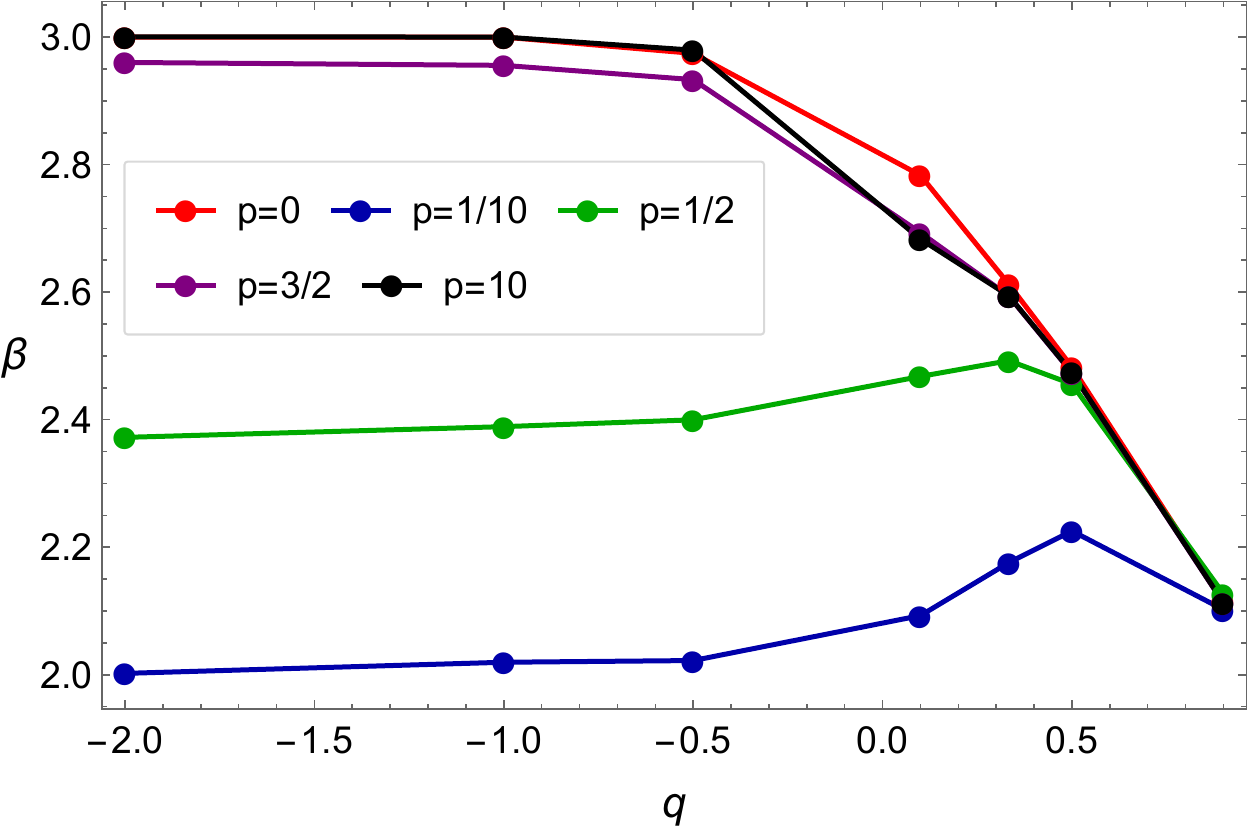}}
  \caption{Behavior of the falloff rate $\beta$ of the self-force as a function of the different values of the redshift parameter $p$ (a)-(e), and (f) shape exponent $q$ for $\sigma=+1$; Taylor's $\beta=3$ result is reproduced via mode-sum approach in (e) where $q=-1$, $p=0$.}
  \label{fig:falloffs}
\end{figure*}

\begin{figure}[htp]
	\centering
	\includegraphics[width=0.95\linewidth]{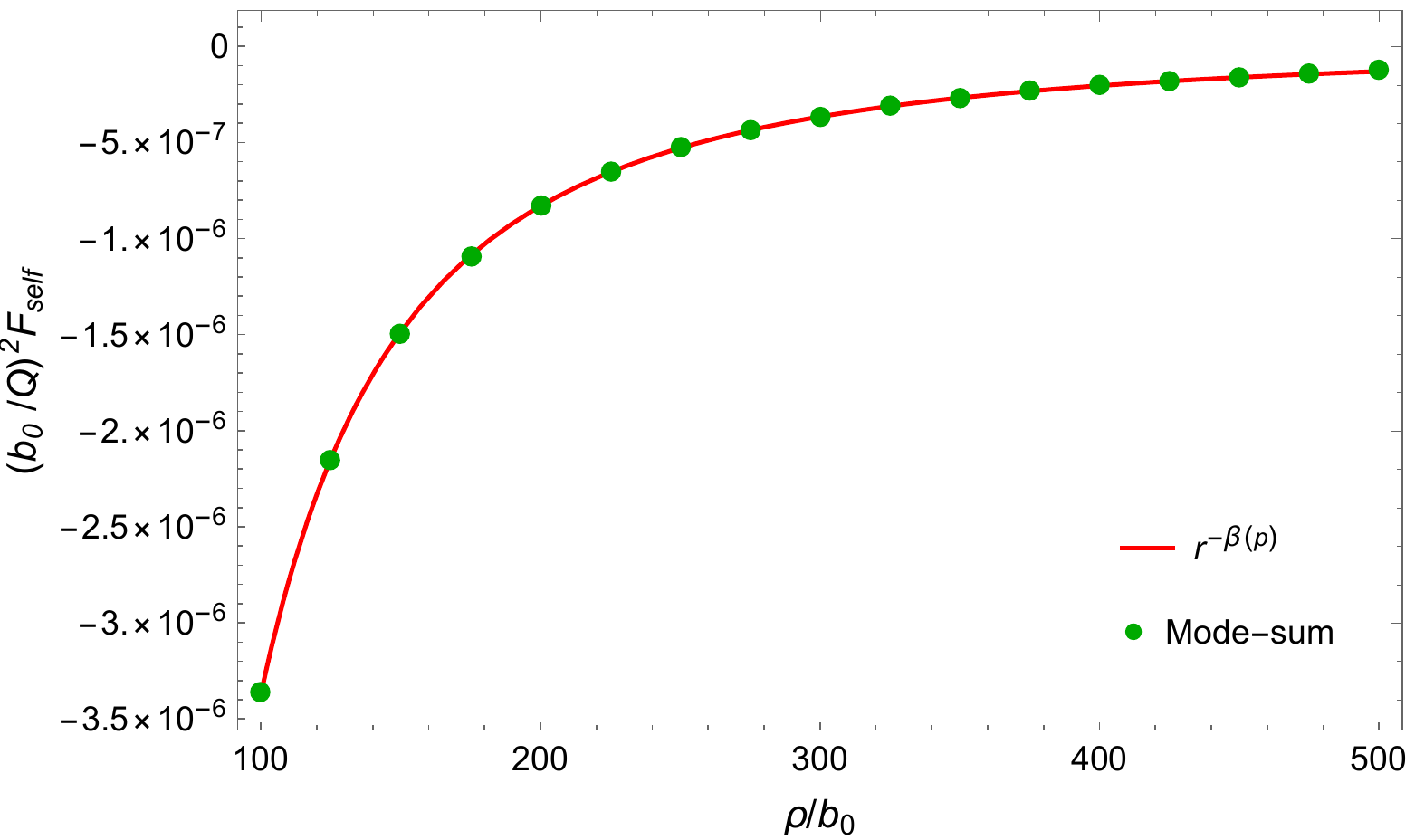}
	\caption{Check of fitting for the power-law distribution against the asymptotic behavior of the self-force for $q=-1$ and $p=1/10$ obtained via mode-sum.}
	\label{fig:fitting}
\end{figure}

\subsection{Asymptotic behavior far from the throat}
In Refs. \cite{de2012probing,de2013probing}, an analytic expression of the self-force was obtained for a static point charge $q$ in the spacetime of a thin shell wormhole. For a certain range of values for the wormhole throat $c$, the self-force can either be repulsive or attractive depending on the position of the point charge. The asymptotic behavior of the self-force computed is given by
\begin{equation}\label{de celis}
    F^{\mathrm{self}} \sim-\frac{Q^{2} c}{2 r^{3}}\left(1-\frac{2 m}{c}\right)+\mathcal{O}\left(r^{-5}\right)
\end{equation}
where $m$ is the mass of the wormhole. Far from the neighborhood of the wormhole, the self-force is always attractive, and it gets infinitely large very near the throat where the curvature diverges. Practically, the leading term in Eq.~\eqref{de celis} can be used to describe the falloff of the self-force numerically computed in this work. In terms of the shape exponent $q$ and redshift parameter $p$, the asymptotic behavior of the self-force in $\rho$ can be approximated by a simple power-law model of the form
\begin{equation}\label{power-law}
    F_{\rho}^{\mathrm{self}}\sim \rho^{-\beta(q, p)}.
\end{equation}
In Fig.~\ref{fig:falloffs}, the asymptotic behavior is plotted in terms of $p$ and $q$. Each point corresponds to a specific $(q, p)$ wormhole and is obtained by fitting the computed $F^{\mathrm{self}}(\rho)$ curve to the power-law model in Eq.~\eqref{power-law}. A straightforward check of the fitting is to examine the behavior of Taylor's result in Eq.~\eqref{taylor} far from the wormhole throat, i.e.,
\begin{equation}
  F_{\rho}^{\mathrm{self}}\sim \rho^{-3}.
\end{equation}
This matches our $\beta\approx 3$ result for the Ellis wormhole $(q=-1, p=0)$ in Fig. \ref{fig:beta5}. For illustration, we have in Fig.~\ref{fig:fitting} a sample fitting of the large-distance behavior of the self-force for the wormhole $(q=-1, p=1/10)$ to the power-law ansatz in Eq.~\eqref{power-law}.

In contrast with other throat profiles as in Figs.~\ref{fig:beta1}-\ref{fig:beta3}, the cases $q=-1/2$ (Fig.~\ref{fig:beta4}) and $q=-1$ (Fig.~\ref{fig:beta5}) exhibit discontinuous behavior in the range $0<p\leq 1$. If we recall the self-force plots for $q=-1$ in Fig.~\ref{fig:tpneg1}, this behavior is to be expected because the self-force undergoes a significant change in its direction and magnitude in response to an increase in $p$. Meanwhile, all of the throat profiles considered manifest very similar behavior, particularly for large values of $p$. Specifically, the falloff rate $\beta(p)$ asymptotes to its corresponding $p=0$ value in the limit as $p\rightarrow\infty$. Recall that in Fig.~\ref{fig:redshift}, the redshift effect $\Delta$ of the wormhole approaches a constant value in $r$ as $p$ increases (see, for instance, $p\geq 5$ when $r>2$). This could provide us with intuition as to why the decay rate $\beta$ of the self-force does not respond significantly with an increase in redshift parameter starting from $p=5$. In fact, this is exactly the case if we take into consideration the redshift function in Eq.~\eqref{minkowski limit}: the limit as $p\rightarrow\infty$ reduces to the ultrastatic case. In terms of the shape exponent $q$, $\beta$ also approaches a constant value given a specific redshift parameter (see Fig.~\ref{fig:sf asymp q}). In addition, notice the overlapping points in Fig.~\ref{fig:sf asymp q} representing negative shape exponents with a large redshift parameter value: the falloff asymptotes to $\beta=3$ in the limit as $p\rightarrow\infty$ and $q\rightarrow-\infty$. This means that the self-force would scale as
\begin{equation}
F_{\rho}^{\mathrm{self}}\sim \rho^{-3},\quad p\rightarrow\infty \quad q\rightarrow-\infty,
\end{equation}
which is the value obtained by Taylor for an ultrastatic traversable wormhole with a shape described by $r=\sqrt[]{\rho^2+b_{0}^2}$, known in the literature as the Ellis wormhole. It is also worth mentioning the $q=9/10$ case to which all points in Fig. \ref{fig:sf asymp q} seem to converge. This means that far from the wormhole throat, the self-force falloff does not change significantly with respect to $p$. In general, the self-force acting on a static scalar charge decays faster for a wormhole with a more negative shape exponent, pictured as having a greater degree of flaring out in the geometry.

\begin{figure}[htp]
	\centering
	\includegraphics[width=0.9\linewidth]{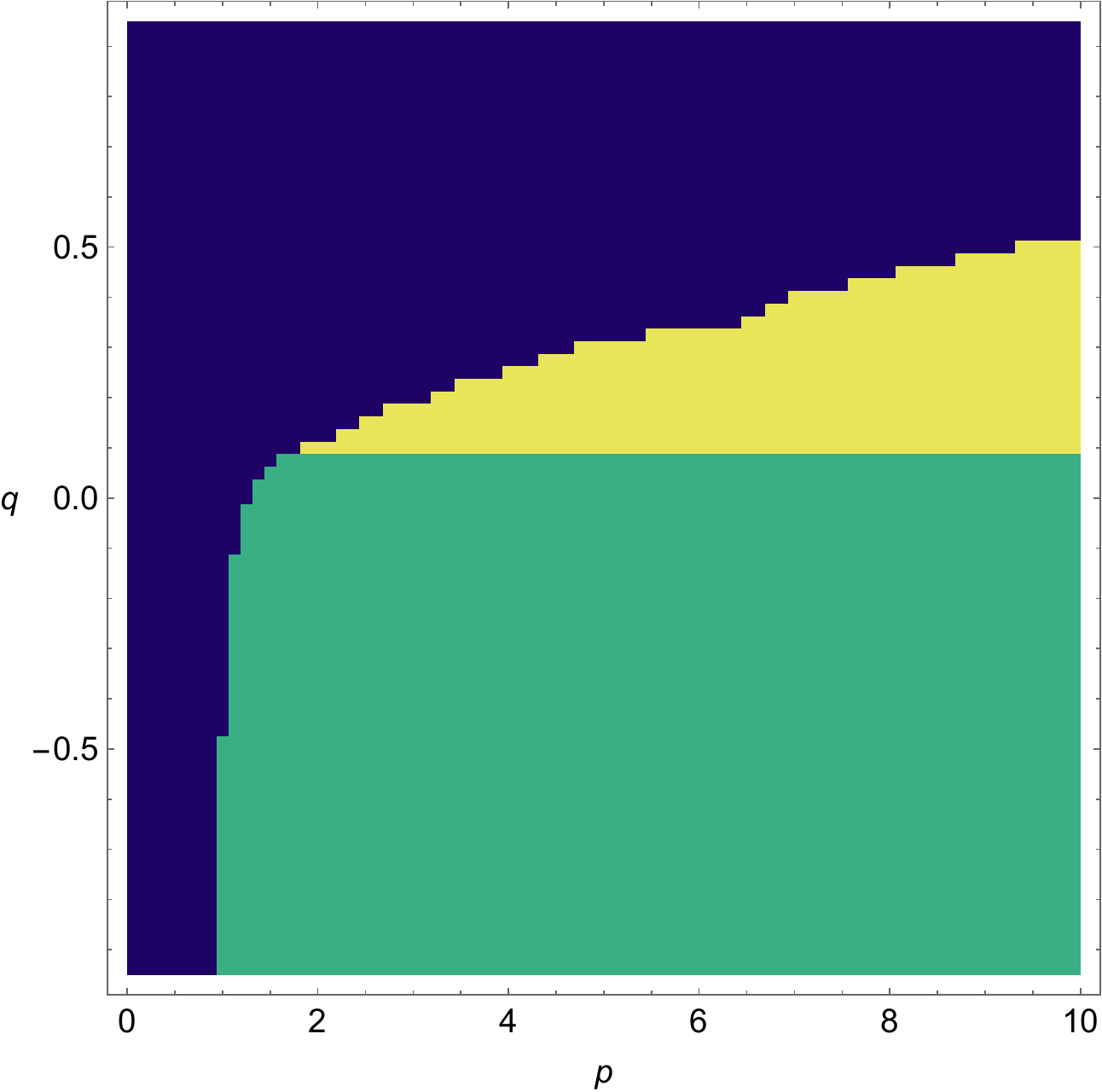}
	\caption{Number of self-force crossings as a function of the wormhole parameters $(p,q)$ where the blue, green, and yellow regions denote zero, one, and two self-force crossings, respectively.}
	\label{fig:sfdiagram}
\end{figure}

\subsection{Number of self-force crossings in terms of the wormhole parameters $(p,q)$}

We present in Fig. \ref{fig:sfdiagram} the number of critical points for which the self-force on a static charge vanishes as a function of the wormhole parameters $(p,q)$. The sign of the self-force is negative near the wormhole throat, and depending on the value of $p$ and $q$, the sign of the self-force changes far from the throat. In the figure above, the blue region represents values of $p$ and $q$ for which no self-force is present. This means that the self-force remains negative far from the wormhole throat. The green region yields one self-force crossing, which signifies that the self-force transitions to being positive far from the throat. Finally, we also observed the presence of two self-force crossings represented by the yellow region, which means that the computed self-force for points far from the throat is negative.

\section{Summary}
\label{sec:Summary}

We have performed a mode-sum calculation of the static scalar self-force in Konoplya–Zhidenko wormholes, extending standard self-force techniques to backgrounds specified numerically for generic values of the geometric parameters. While the mode-sum framework itself is well established, our implementation uses numerically generated metric functions, enabling a systematic exploration of the two-dimensional parameter space $(p,q)$ controlling throat geometry and gravitational redshift.

As a consistency check, we verified that our numerical results reproduce the known analytical solution for the Ellis wormhole, for which the self-force is everywhere repulsive. This agreement was achieved through careful implementation of the mode-sum regularization procedure and systematic monitoring of the large-$\ell$ falloff of the multipole contributions, which provided a stringent test of the calculation.

Our analysis reveals several qualitative features not previously seen in the extant literature. Most notably, the self-force is generally not of fixed sign: depending on $(p,q)$ and the particle’s location, it may reverse direction, with up to two radial positions at which it vanishes. By comparing ultrastatic configurations with their $\Phi \neq 0$ counterparts, we demonstrated that these zero crossings are closely associated with the presence of gravitational redshift (nonzero $p$), highlighting the role of tidal effects in shaping the force.

We further showed that increasing $p$ enhances the overall magnitude of the self-force, while increasing $|q|$ strengthens geometric flaring and modifies its radial decay. In particular, for large $p$, more strongly flared wormholes exhibit slower-than-$r^{-3}$ falloff, in contrast with the behavior typical of isolated-body spacetimes. In the joint limit $p \to \infty$ and $q \to -\infty$, the asymptotic decay approaches that of the Ellis wormhole, thereby providing a unifying endpoint for the parameter space.

Taken together, these results demonstrate that the scalar self-force provides a sensitive probe of wormhole geometry, encoding both redshift and throat structure in its sign and radial scaling.

\section*{Acknowledgements}
This research is supported by the University of the Philippines OVPAA through Grant No.~OVPAA-BPhD-2016-13 and the DOST-SEI through the ASTHRDP scholarship.

\appendix
\section{General throat profile}\label{appendix a}
For an arbitrary shape exponent $q$, we have obtained a general expression of the coordinate transformation $r\rightarrow r(\rho)$. We can simplify the integral in Eq.~\eqref{properdistance} by performing the following change of variables,
\begin{align}
u=&\left(\frac{b_0}{\overline{r}}\right)^{1-q}, \qquad \alpha=\frac{1}{q-1}-1, \qquad\eta=\left(\frac{b_0}{r}\right)^{1-q}
\end{align}
Hence, Eq.~\eqref{properdistance} simplifies to
\begin{equation}\label{int1}
    \rho=\frac{b_0}{q-1}\int_1^\eta u^\alpha(1-u)^{-1/2}du
\end{equation}
from which we recall the integral representation of the incomplete beta function given by
\begin{equation}
B(x;a,c)=\int_0^x w^{a-1}(1-w)^{c-1}dw, \label{beta}
\end{equation}
where $0\leq x\leq 1$ and $a,c>0$. Again, we make another change of variable $v=1-u$ to express Eq.~\eqref{int1} in the form of Eq.~\eqref{beta}
\begin{align}
\rho(r)=&-\frac{b_0}{q-1}\int_0^{1-\eta}(1-v)^\alpha v^{-1/2}dv\\
=&\frac{b_0}{1-q}B\left(1-\left(\frac{b_0}{r}\right)^{1-q};\frac{1}{2},\frac{1}{q-1}\right)\label{beta1}
\end{align}
Note that $1/(q-1)<0$ for $-\infty<q<1$, hence Eq. \eqref{beta1} does not hold for our range of $q$. But we can, in principle, extend the domain of the incomplete beta function for negative values of $1/(q-1)$ by means of analytic continuation. First, we write the binomial expansion of the term in parentheses in Eq.~\eqref{beta}
\begin{align}
(1-w)^{c-1}=&\sum_{n=0}^\infty \binom{c-1}{n}(1)^{c-1-n}(-w)^n\\
=&\sum_{n=0}^\infty\frac{(c-1)_n}{n!}(-w)^n\label{binomial}
\end{align}
where $(.)_n$ is the Pochhammer symbol. Using Eq.~\eqref{binomial} in Eq.~\eqref{beta1} gives
\begin{align}
\rho(r)=&\frac{b_0}{1-q}\int_0^x\sum_{n=0}^\infty\frac{(c-1)_n}{n!}(-w)^n w^{a-1}dw\\
=&\frac{b_0}{1-q}\int_0^x\sum_{n=0}^\infty\frac{(c-1)_n}{n!}(-1)^n w^{a+n-1}dw\nonumber\\
=&\frac{b_0}{1-q}\sum_{n=0}^\infty\frac{(c-1)_n}{n!}\frac{(-1)^n}{a+n} x^{a+n}\nonumber\\
=&\frac{b_0}{1-q}\frac{x^a}{a}\sum_{n=0}^\infty(-1)^n\frac{(c-1)_n}{n!}\frac{(a)_n}{(a+1)_n} x^n\nonumber\\
=&\frac{b_0}{1-q}\frac{x^a}{a}\sum_{n=0}^\infty\frac{(a)_n(1-c)_n}{(a+1)_n}\frac{x^n}{n!}. \label{eq98}
\end{align}

Notice that the sum in Eq.~\eqref{eq98} represents the hypergeometric function in terms of the Pochhammer symbol
\begin{equation}
_2F_1(k,l;m;x)=\sum_{n=0}^\infty\frac{(k)_n(l)_n}{(m)_n}\frac{x^n}{n!}
\end{equation}
where $k=a=1/2$, $l=1-c=1-1/(q-1)$, and $m=a+1=3/2$.
Hence, we obtain the analytic continuation of the incomplete beta function in terms of the more generic Gauss hypergeometric function given by
\begin{align}\label{coord transform}
    \rho ( r ) = \,&\frac { 2 b _ { 0 } } { 1 - q }\, \sqrt { 1 - \left( \frac { b _ { 0 } } { r } \right) ^ { 1 - q } } \nonumber\\
    &\times \,_ { 2 } F _ { 1 } \left( \frac { 1 } { 2 } , \frac { q - 2 } { q - 1 } ; \frac { 3 } { 2 } ; 1 - \left( \frac { b _ { 0 } } { r } \right) ^ { 1 - q } \right).
\end{align}
Inverting Eq.~\eqref{coord transform} gives the wormhole throat profile $r(\rho)$. Unfortunately, an exact expression for the throat profile cannot be obtained since, in general, Eq.~\eqref{coord transform} cannot be inverted analytically.

\section{Scalar field equation}\label{appendix b}
The metric expressed in terms of the proper radial coordinate $\rho$ is given in Eq.~\eqref{metric1} with determinant $g=-e^{2\Phi} r^4\sin^2\theta$, where $\Phi=\Phi(\rho)$ and $r=r(\rho)$. The field $\Psi$ generated by the scalar charge is a solution to the following wave equation
\begin{equation}
(\Box -m^2-\xi R)\Psi=-4\pi j
\end{equation}
where $m$ is the mass of the scalar field with coupling $\xi$ to the curvature scalar $R$, $j$ denotes the charge density
\begin{equation}
j(x)=Q\int \delta^{(4)}(x-x_s(\tau))\frac{d\tau}{\sqrt[]{-g}},
\end{equation}
while the d'Alembertian wave operator is explicitly given by
\begin{align}
\square=\frac{1}{\sqrt[]{-g}}\partial_{\mu}\Big(\sqrt[]{-g}g^{\mu\nu}\partial_{\nu}\Big)
\end{align}
where $\partial_\mu$ denotes a partial differentiation with respect to coordinate $\mu$. For a massless $(m=0)$ scalar field with minimal coupling ($\xi=0$) to the curvature scalar $R$, the wave equation simply becomes
\begin{equation}\label{wave_eq1}
\Box\Psi=-4\pi j.
\end{equation}
The wave operator is explicitly evaluated as
\begin{align}\label{wave operator}
\square=&\frac{1}{\sqrt[]{-g}}\Big[-\partial_t\left(\sqrt[]{-g}g^{tt}\partial_t\right)+\partial_{\rho}\left(\sqrt[]{-g}g^{\rho\rho}\partial_\rho\right)\nonumber\\
&+\partial_{\theta}\left(\sqrt[]{-g}g^{\theta\theta}\partial_\theta\right)+\partial_{\phi}\left(\sqrt[]{-g}g^{\phi\phi}\partial_\phi\right)\Big]\nonumber\\
=&\frac{1}{e^{\Phi}r^2\sin\theta}\Big[-\partial_t\left( e^{\Phi}r^2\sin\theta \cdot \, e^{-2\Phi}\partial_t\right)\nonumber\\
&+\partial_{\rho}\left(e^{\Phi}r^2\sin\theta\partial_{\rho}\right)+\partial_{\theta}\left(e^{\Phi}r^2\sin\theta \cdot \, r^{-2}\partial_{\theta}\right)\nonumber\\
&+\partial_{\phi}\left(e^{\Phi}r^2\sin\theta \cdot \,(r\sin\theta)^{-2}\partial_{\phi}\right)\Big]\nonumber\\
=&-\frac{1}{e^{2\Phi}}\partial_t^2+\frac{1}{e^{\Phi}r^2}\partial_{\rho}\left(e^{\Phi}r^2\partial_{\rho}\right)+\frac{1}{r^2\sin\theta}\partial_{\theta}\left(\sin\theta \partial_{\theta}\right)\nonumber\\
&+\frac{1}{r^2\sin^2\theta}\partial_{\phi}^2
\end{align}

For the charge at rest at point $x_s$ according to observers moving on integral curves of the Killing vector $\partial/\partial t$, the wave equation reduces to the three-dimensional Helmholtz equation
\begin{align}
\frac{1}{e^{\Phi}r^2}\partial_{\rho}&\left(e^{\Phi}r^2\partial_{\rho}\Psi\right)+\frac{1}{r^2\sin\theta}\partial_{\theta}\left(\sin\theta \partial_{\theta}\Psi\right)\nonumber\\
&+\frac{1}{r^2\sin^2\theta}\partial_{\phi}^2\Psi=-4\pi j
\end{align}
The source density becomes
\begin{align}
j(x)=&\,Q\int \delta^{(4)}(x-x_s(\tau))\frac{d\tau}{\sqrt[]{-g}}\\
=&\frac{Q}{e^{\Phi(r)}r^2 \sin\theta}\delta^{(3)}(x-x_s)\int \delta(t-t_s(\tau))d\tau\\
=&\frac{Q}{e^{\Phi(r)}r^2 \sin\theta}\delta^{(3)}(x-x_s)\int \frac{\delta(t-t_s(\tau))}{e^{-\Phi(r)}}dt\\
=&\frac{Q}{r^2 \sin\theta}\delta^{(3)}(x-x_s).
\end{align}

In order to integrate the field equation, we implement the multipole decomposition of the potential
\begin{equation}\label{field deco}
\Psi(\rho,\theta,\phi)=\sum_{lm}\psi_{lm}(\rho)Y_{lm}(\theta,\phi).
\end{equation}
Similarly, the source is decomposed into
\begin{equation}\label{source deco}
j(x)=\frac{Q\,\delta(\rho-\rho_s)}{r^2}\sum_{lm}Y^*_{lm}(\theta_s,\phi_s)Y_{lm}(\theta,\phi)
\end{equation}
in which $(\rho_s,\theta_s,\phi_s)$ represent the spherical coordinates of the particle's position $x'$, where we have used the spherical harmonic closure relation given by
\begin{align}\label{closure rel}
    \sum _ { l = 0 } ^ { \infty } \sum _ { m = - l } ^ { l } &Y _ { lm } \left( \theta _ { 1 } , \phi _ { 1 } \right) { Y } _ { lm } ^ { * } \left( \theta _ { 2 } , \phi _ { 2 } \right) \nonumber\\
    &= \frac { 1 } { \sin \theta _ { 1 } } \delta \left( \theta _ { 1 } - \theta _ { 2 } \right) \delta \left( \phi _ { 1 } - \phi _ { 2 } \right).
\end{align} 
Without loss of generality, we may place the particle along the polar axis $(\theta_s=0)$ and exploit the property of spherical-harmonic functions
\begin{equation}\label{identity}
Y_{lm}(0,\phi)=\sqrt[]{\frac{2l+1}{4\pi}}\delta_{m,0}.
\end{equation}
Substitution of Eqs.~\eqref{field deco}, \eqref{source deco}, \eqref{closure rel}, \eqref{identity} within the field equation then produces
\begin{align}
 r^2\psi''&+\left(\Phi'+\frac{2r'}{r} \right)r^2\psi'-l(l+1)\psi\nonumber\\
 &=-4\pi Q\,\sqrt[]{\frac{2l+1}{4\pi}}\delta(\rho-\rho_s)
\end{align}
where $\psi=\psi_{l0}(\rho)$ with the prime here denoting differentiation on $\rho$. The modes with spherical harmonic index $m\neq 0$ necessarily vanish. 

We perform nondimensionalization with the transformation
\begin{equation}
y=\rho/b_0
\end{equation}
as well as introduce $\tilde{x}=r/b_0$. The radial equation takes a simpler form
\begin{align}\label{eqn in y}
\tilde{x}^{2}\psi''(y)+&\left(2\tilde{x}-\frac{\sigma\,p}{\tilde{x}^{p-1}}\right)\,\tilde{x}'\psi'(y)-l(l+1)\psi(y)\nonumber\\
&=-4\pi Q\,\sqrt[]{\frac{2l+1}{4\pi}}\frac{\delta(y-y_s)}{b_0}
\end{align}
with prime denoting differentiation on $y$.
The general solution may be written as a sum of two linearly independent solutions 
\begin{equation}\label{ansatz}
\psi=c_{1}\,\psi_{<}\,\Theta(y_s-y)+c_{2}\,\psi_{>}\,\Theta(y-y_s)
\end{equation}
of the homogeneous equation
\begin{equation}\label{homo-eq}
    \tilde{x}^{2}\psi''(y)+\left(2\tilde{x}-\frac{\sigma\,p}{\tilde{x}^{p-1}}\right)\,\tilde{x}'\psi'(y)-l(l+1)\psi(y)=0
\end{equation}
where
\begin{equation}
\psi= \begin{cases}c_{1} \psi_{<}, & \text {if } y<y_{s} \\ c_{2} \psi_{>}, & \text {if } y>y_{s}\end{cases}
\end{equation}
and $\Theta$ is the Heaviside step function. Substituting Eq.~\eqref{ansatz} into \eqref{homo-eq} yields the following expressions for the solution $\psi$ and its derivatives
\begin{align}
\psi(y)=&\,c_1\psi_<\Theta(y_s-y)+c_2\psi_>\Theta(y-y_s)\label{psi}\\
\psi'(y)=&\,c_2\psi_>'\Theta(y-y_s)+c_2\psi_{>}(y_s)\nonumber\\
&+c_1\psi_<'\Theta(y_s-y)-c_1\psi_<'(y_s)\label{psi prime}\\
\psi''(y)=&\,c_1\psi_<''\Theta(y_s-y)+c_2\psi_>''\Theta(y_s-y)\nonumber\\
&+(c_2\psi_>'-c_1\psi_<')\delta(y-y_s)\label{psi double prime}
\end{align}
from which we obtain the jump condition at the location of the charge $y=y_s$. Substituting Eqs.~\eqref{psi}, \eqref{psi prime}, and \eqref{psi double prime} into Eq.~\eqref{eqn in y} and matching terms containing the delta function, we arrive at the following junction condition
\begin{equation}\label{jump condition}
\left[c_2\psi_>'-c_1\psi_<'\right]_{y=y_s}=-\frac{Q}{b_{0}\tilde{x}_{s}^2}\sqrt[]{4\pi(2l+1)}
\end{equation}
where $\tilde{x}_s=\tilde{x}(y_s)$. At the boundary, the solutions must be continuous
\begin{equation}\label{continuity}
c_1\psi_<(y_s)=c_2\psi_>(y_s).
\end{equation}
The amplitude $c_1$ can be obtained by combining Eqs.~\eqref{jump condition} and \eqref{continuity} as
\begin{equation}\label{c1}
c_1=-\frac{Q \, \sqrt[]{4\pi(2l+1)}}{b_{0}\tilde{x}_{s}^2}\left(\frac{\psi_>}{\psi_<\psi_>'-\psi_<'\psi_>}\right),
\end{equation}
while $c_2$ is taken from the continuity condition as
\begin{equation}\label{c2}
    c_2=-\frac{Q \, \sqrt[]{4\pi(2l+1)}}{b_{0}\tilde{x}_{s}^2}\left(\frac{\psi_<}{\psi_<\psi_>'-\psi_<'\psi_>}\right)
\end{equation}
where $\psi_<$ and $\psi_>$ are to be evaluated at the particle location $y=y_s$.

\section{Asymptotic solution}\label{appendix c}
The homogeneous part of the wave equation for a static scalar charge in a wormhole with throat profile $ \tilde{x}(y)=\sqrt{y^2+1} $ and arbitrary redshift parameter $p=n/m$ is obtained as
\begin{align}\label{odey ellis}
   \left(y^2+1\right) \psi''(y)&+ \, y \left(2-p \left(y^2+1\right)^{-p/2}\right) \psi'(y)\nonumber\\
   &-l (l+1) \psi(y)=0.
\end{align}
The presence of a non-zero $p$ makes the coefficients of the differential equation non-polynomial. To be able to apply the Frobenius method, we perform the following change of variable
\begin{equation}
z=\left(y^2+1\right)^{1/2m}
\end{equation}
to obtain polynomial coefficients
\begin{align}\label{odez ellis}
   z^{2}\left(1-z^{-2m}\right) \psi''(z)+& \,z^{1-2m-n}\left(z^{2m}\left((1+m)z^{n}-n\right)\right.\nonumber\\
   \left.-z^{n}+n\right) \psi'(z)&-m^2 l (l+1) \psi(z)=0
\end{align}
where $n,m\in\mathbb{N}$ and prime now denotes differentiation with respect to $z$. By inserting the ansatz
\begin{equation}\label{asympz}
    \psi(z)=\dfrac{1}{z^k}\sum_{i=0}^{\infty}\dfrac{a_i}{z^i}
\end{equation}
into Eq.~\eqref{odez ellis}, an asymptotic solution for $\psi(z)$ as $z\rightarrow\infty$ is obtained. Employing the Frobenius method, we find the indicial exponent to be $k=m(l+1)$ and the recurrence relation for the coefficients $a_i$ is derived as  
\begin{align}
a_{i}=&\left(i (i+2 l m+m)\right)^{-1}\left[(i+(l-1) m)^2 a_{i-2 m}\right.\nonumber\\
&-n (i+l m+m-n) a_{i-n}\nonumber\\
&\left.+n (i+(l-1) m-n) a_{i-2 m-n}\right].
\end{align}

\bibliographystyle{apsrev4-1}
\bibliography{bibfile}

\end{document}